\documentclass[12pt]{iopart} 
\usepackage{graphicx} 
\usepackage{amssymb}
\begin{document}

\title{  Magnetic and orbital correlations in a two-site molecule }

\author{ Marcin Raczkowski$^{1,2}$,  
         Raymond Fr\'esard$^2$ and 
	 Andrzej M Ole\'s$^1$ }

\address{$1$ Marian Smoluchowski Institute of Physics,
             Jagellonian University, \\   
             Reymonta 4, PL-30059 Krak\'ow, Poland }

\address{$2$ Laboratoire CRISMAT, UMR CNRS--ENSICAEN(ISMRA) 6508, \\  
             6 Bld. du Mar\'echal Juin, F-14050 Caen, France}  

\ead{M.Raczkowski@if.uj.edu.pl, 
     Raymond.Fresard@ensicaen.fr and 
     A.M.Oles@fkf.mpg.de}

\begin{abstract}
We analyze the role of orbital degeneracy in possible magnetic and 
orbital instabilities by solving exactly a two-site molecule with two 
orbitals of either $e_g$ or $t_{2g}$ symmetry at quarter-filling. As
a generic feature of both models one finds that the spin and orbital
correlations have opposite signs in the low temperature regime when 
the orbitals are degenerate, in agreement with the Goodenough-Kanamori 
rules. While Hund's exchange coupling $J_H$ induces ferromagnetic spin 
correlations in both models, it is more efficient for $t_{2g}$ orbitals
where the orbital quantum number is conserved along the hopping 
processes. We show that the ground state and finite temperature 
properties may change even qualitatively with increasing Coulomb 
interaction when the crystal field splitting of the two orbitals 
is finite, and the Goodenough-Kanamori rules may not be followed.\\ \\
{\it Published in J. Phys.: Condens. Matter {\bf 18}, 7449-7469 (2006).}
\end{abstract}

\pacs{71.10.Fd, 71.27.+a, 75.10.Lp, 75.30.Et}  


\maketitle

\section{Introduction}

The Hubbard model has been employed for a long time as the conceptually 
simplest model which might explain metallic ferromagnetism of 
itinerant electrons \cite{Kan63,Hub63} and localization due to Coulomb
repulsion \cite{Gut63}. Unlike initially expected, this model does not 
easily yield ferromagnetism on the hypercubic lattice and some
additional conditions have to be satisfied to stabilize a ferromagnetic 
(FM) phase. As one of very few exact result in this many-body problem, 
Nagaoka established long ago \cite{Nag66} that in the limit of infinite 
local Coulomb interaction $U=\infty$, a single hole doped into a 
half-filled system leads to the FM ground state. There are several 
indications that this remains valid for a finite concentration of holes 
\cite{vdL91,Han97}. Nevertheless the ground state of the Hubbard 
model on the square lattice may only be FM for values of $U$ far larger
than the bandwidth \cite{Mol93b}. Ferromagnetism may be also promoted 
by a particular lattice or band structure \cite{Han97}, or even by 
disorder \cite{Byc03}. Lieb first showed that a half-filled flat band 
induces a net magnetization \cite{Lie89}. Furthermore, Hirsch and others 
have focused on the effect of additional off-diagonal matrix elements of 
the Coulomb interaction \cite{Hir89}.

A major step towards understanding the physics of metallic 
ferromagnetism in transition metals such as Fe, Co, and Ni, was the 
suggestion that orbital degeneracy might play a crucial role. It was 
first pointed out by Slater, Statz, and Koster \cite{Sla53} and then 
stressed by van Vleck \cite{Vle53} that in the presence of degenerate 
orbitals, Hund's exchange coupling $J_H$ favours local triplet spin 
configurations of two electrons occupying different orbitals. This has 
important consequences for ferromagnetism as local moments are formed 
and thus $J_H$ helps to stabilize magnetic phases, including the FM one 
\cite{Hel98}. Roth examined
the doubly degenerate model at quarter-filling in three dimensions 
\cite{Rot66}. She found that the ground state is a spin triplet and 
orbital singlet for two sites, i.e., the system forms an orbital 
superlattice structure in which two sublattices have different orbitals 
occupied by electrons at each of them. This interrelation between 
staggered orbital order and spin ferromagnetism was next emphasized by 
Kugel and Khomskii \cite{Kug73} who derived an effective strong coupling 
Hamiltonian with coupled spin and orbital degrees of freedom, extended 
further by Cyrot and Lyon-Caen \cite{Cyr75} by on-site pair hopping, and 
by Inagaki \cite{Ina75}. These seminal papers started a new field --- 
spin-orbital physics in correlated transition metal oxides \cite{Mae04}, 
where superexchange models derived in the strong coupling regime provide 
a theoretical background for understanding both magnetic and optical
properties \cite{Ole05}. Numerous spin-orbital models have been derived
in the regime of large Coulomb interactions, both with $e_g$ 
\cite{Ole00} and with $t_{2g}$ \cite{Kha00} orbital degrees of freedom 
and are currently investigated.

While systems of higher dimensionality are clearly the ones of most  
interest \cite{iso,anizo}, significant insight into the complementary 
behaviour of spin and orbital degrees of freedom was obtained in a 
one-dimensional (1D) model. Indeed, many of essential features of such 
1D systems were established by quantum Monte Carlo simulations 
\cite{Gil87}, Exact Diagonalization (ED) studies 
\cite{Spa79,Kue97,Hir97}, the combination of these two approaches 
\cite{Dag04}, and  Density Matrix Renormalization Group 
(DMRG) method \cite{Sak02}. Spin, charge, and orbital correlations in 
the 1D Hubbard model with $t_{2g}$ orbitals at several densities have 
been also examined in a very recent study by Xavier \textit{et al.} 
\cite{Xav06}, which combines DMRG and the Lanczos technique.  

Unfortunately, in a vast majority of studies, a conceptually simplified
model, i.e., the degenerate Hubbard model with equivalent orbitals 
\cite{iso} or at best with different bandwidth obtained with diagonal 
hopping \cite{anizo}, has been investigated so far. However, it has 
been recently shown within dynamical mean field theory that a finite 
on-site hybridization between orbitals enhances the charge and orbital 
fluctuations and plays a significant role especially for the 
orbital-selective Mott transition \cite{Kog05}. In fact, a proper 
description of a system with $e_g$ orbitals involves a more complex 
kinetic energy Hamiltonian as in this case the orbital flavour during 
the intersite hopping is not conserved, and this is likely to lead to 
partial orbital polarization which is expected  to strongly modify the 
magnetic instabilities \cite{Fre05,Rac05}, resulting in a rich phase 
diagram. In addition only little is known in analytic form about these 
models, besides expansions in $t$ around the atomic limit which result 
into the celebrated Goodenough-Kanamori rules \cite{Goode,Ole06} and 
spin-orbital models. In particular cases, such as half-filling in 
two-band models, they corroborate the results obtained from weak 
coupling approaches, and therefore gaining further qualitative insight 
into the corresponding problems is unlikely. In contrast, at quarter 
filling, weak coupling approaches are at odd with the strong coupling 
expansions \cite{Fre05}, and the situation remains controversial. 
Fortunately this situation can be studied rigorously on the analytical 
level, both for $e_g$ and $t_{2g}$ orbitals, in a two-site molecule. 

The paper is organized as follows. Firstly, we introduce a 
\emph{realistic} model with two $e_g$ orbitals and compared it with the 
one for $t_{2g}$ orbitals in \sref{sec:Model}. Secondly, the solutions 
for both models are given and compared with each other in 
\sref{sec:deg}. Thereby, we identify characteristic differences in the 
behaviour of both orbital degrees of freedom \cite{vdB04}. Furthermore, 
in \sref{sec:deg} we verify whether the phenomenon of a complementary 
behaviour of the spin and orbital flavours is also a characteristic 
feature of this particular case by analyzing spin and pseudospin 
correlation functions. Comparisons to strong coupling expansions are 
presented as well. Next, in \sref{sec:nondeg} we investigate the 
influence of finite crystal field splitting on the ground state 
properties. Finally, \sref{sec:sum} summarizes the paper and gives 
general conclusions.

\section{Model Hamiltonian for $e_g$ and for $t_{2g}$ orbitals}
\label{sec:Model}

The magnetic and orbital instabilities within the $e_g$ band become
especially relevant in the context of doped nickelates
La$_{2-x}$Sr$_x$NiO$_4$ (LSNO), where interesting novel phases including  
the stripe order were discovered \cite{Sac95}. Even though doped 
nickelate LSNO is isostructural with its cuprate counterpart 
La$_{2-x}$Sr$_x$CuO$_4$, its electronic degrees of freedom are 
more complicated. In fact, a realistic Hamiltonian for LSNO must 
contain, besides the $x^{2}-y^{2}$ orbital states deciding about the 
properties of the cuprates, also the $3z^{2}-r^{2}$ orbital states, 
so as to account for the actual filling with two holes and for the 
high-spin state ($S=1$) in the stoichiometric compound. Such a model of 
interacting $e_g$ electrons in ($a,b$) plane may be written as follows,
\begin{equation}
{\cal H}=H_{\mathnormal kin} + H_{\mathnormal int} + H_{\mathnormal cf},
\label{eq:H_deg}
\end{equation}
with two orbital flavours: $|x\rangle\sim x^{2}-y^{2}$ and
$|z\rangle\sim 3z^{2}-r^{2}$ forming a basis in the orbital space. 
The kinetic energy is described by,
\begin{equation}
H_{\mathnormal kin}= \sum_{\langle ij\rangle}\sum_{\alpha\beta\sigma}
    t^{\alpha\beta}_{ij} c^{\dag}_{i\alpha\sigma}c^{}_{j\beta\sigma},
    \qquad t^{\alpha\beta}_{ij}=-\frac{t}{4}\left(\begin{array}{cc} 3
	& \pm\sqrt{3} \\ \pm\sqrt{3} &  1
\end{array} \right),
\label{eq:H_kin}
\end{equation}
where $t$ stands for an effective $(dd\sigma)$ hopping matrix element 
due to the hybridization with oxygen orbitals on Ni$-$O$-$Ni bonds, 
and the off-diagonal hopping $t^{xz}_{ij}$ along $a$ and $b$ axis 
depends on the phase of the $|x\rangle$ orbital along the considered 
cubic direction. 

For our purpose it is most convenient to consider an $e_g$ orbital basis
consisting of a directional orbital $|\zeta\rangle$ along the molecular 
bond and a planar orbital $|\xi\rangle$ orthogonal to $|\zeta\rangle$
\cite{Dag04}. Pairs of such orthogonal orbitals defining a new basis  
might be obtained by the following transformation of the original 
orbital basis $\{|x\rangle,|z\rangle\}$,
\begin{equation}
\left(\begin{array}{c} 
|\zeta\rangle \\ 
|\xi\rangle
\end{array}\right) =
\left(\begin{array}{rc}
 \cos\case{\theta}{2}  & \sin\case{\theta}{2} \vspace{0.1cm} \\
-\sin\case{\theta}{2}  & \cos\case{\theta}{2}
\end{array}\right) 
\left(\begin{array}{c} 
|z\rangle \\ 
|x\rangle
\end{array}\right),
\label{eq:rot}
\end{equation}
with the angle $\theta=\pm2\pi/3$ depending on whether one considers 
the bond along $a$ or $b$ axis. This rotation leads to a simple 
diagonal form of the hopping matrix,
\begin{equation}
t^{\zeta\xi}_{ij}=-t 
\left(\begin{array}{cc} 
1  & 0 \\ 
0  & 0
\end{array}\right),
\label{eq:H_eg}
\end{equation}
allowing only for intersite transitions between two directional
$|\zeta\rangle$ orbitals along each bond \cite{Fei05}. 
We compare this case with a frequently studied 
degenerate Hubbard model with equivalent orbitals. Such a model 
describes the dynamics of two active $t_{2g}$ orbitals, e.g., 
for a molecular bond along $a$ axis, $|\zeta\rangle\sim|xz\rangle$ and 
$|\xi\rangle\sim|xy\rangle$, so that the hopping matrix is diagonal,
\begin{equation}
t^{\zeta\xi}_{ij}=-t 
\left(\begin{array}{cc} 
1   & 0 \\ 
0   & 1
\end{array}\right).
\label{eq:h_t2g}
\end{equation}
%


The electron-electron interactions are described by 
the on-site terms, which we write in the following form \cite{Ole83},
\begin{eqnarray}
H_{\mathnormal{int}} &= 
    U\sum_{i}\bigl( n^{}_{ix\uparrow}n^{}_{ix\downarrow} +
    n^{}_{iz\uparrow}n^{}_{iz\downarrow}\bigr ) 
    +\bigl(U-\case{5}{2}J_H\bigr)\sum_{i}n^{}_{ix}n^{}_{iz} \nonumber \\ 
  &-2J_H\sum_{i}\textbf{S}_{ix}\cdot\textbf{S}_{iz} 
   +J_H\sum_{i}\bigl
   (c^{\dag}_{ix\uparrow}c^{\dag}_{ix\downarrow}
    c^{}_{iz\downarrow}c^{}_{iz\uparrow} +
    c^{\dag}_{iz\uparrow}c^{\dag}_{iz\downarrow}
    c^{}_{ix\downarrow}c^{}_{ix\uparrow} \bigr).
\label{eq:H_int}
\end{eqnarray}
Here $U$ and $J_H$ stand for the intraorbital Coulomb and Hund's 
exchange elements, whereas $n_{i\alpha}=\sum_{\sigma}n_{i\alpha\sigma}$  
is the electron density at site $i$ in $\alpha=x,z$ orbital state. 

The last term $H_{\mathnormal cf}$ stands for the uniform crystal-field 
splitting between $|x\rangle$ and $|z\rangle$ orbitals along the $c$ 
axis,
\begin{equation}
H_{\mathnormal cf} = \case{1}{2}E_{0}\sum_{i\sigma}
(n_{ix\sigma}-n_{iz\sigma}).
\label{eq:H_cf}
\end{equation}
The splitting of two $e_g$ orbitals occurs due to the tetragonal
Jahn-Teller distortion of the NiO$_6$ octahedron. In La$_2$NiO$_4$, 
however, the octahedron, with the Ni$-$O$-$Ni bond lengths being 1.95 
(2.26) \AA\ in-plane (out-of-plane) \cite{Jor89}, respectively, is much 
less distorted than the CuO$_6$ octahedron with 1.89 and 2.43 \AA\ bond
lengths in La$_2$CuO$_4$ \cite{Rad94}, which reflects the difference in 
electron filling. In what follows we consider only a realistic positive 
$E_0$ favouring, due to elongated octahedra, the $|z\rangle$ occupancy 
over the $|x\rangle$ occupancy by the $e_g$ electrons in doped 
compounds.

\section{Physical properties at orbital degeneracy}
\label{sec:deg}

\subsection{Classification of eigenstates using the symmetry properties}
\label{subsec:sym}

In this section we present an exact solution of a two-site molecule at
quarter filling with two degenerate orbitals of either $e_g$ or $t_{2g}$ 
symmetry. Although it is straightforward to solve the present problem 
numerically, it is more instructive to find the solution analytically. 
Along this process several important aspects will be clarified.

While the total spin operator  commutes with the Hamiltonian and its
$z$-component can be used to label the eigenstates this does not hold
for the total orbital pseudospin operator  
$\textbf{T}=\sum_{i}\textbf{T}_{i}$, where its three components are 
given by,
\begin{equation}
\textbf{T}_{i}=\Bigl\{{\mathcal T}_{i}^{+},{\mathcal T}_{i}^{-},
{\mathcal T}_{i}^{\zeta}\Bigr\}= \Bigl\{
\sum_{\sigma}c^{\dag}_{i\xi\sigma}c^{}_{i\zeta\sigma},
\sum_{\sigma}c^{\dag}_{i\zeta\sigma}c^{}_{i\xi\sigma},
\case{1}{2}\sum_{\sigma}( n_{i\xi\sigma}-n_{i\zeta\sigma} )\Bigr\}.
\label{eq:pseudospin}
\end{equation}
In order to distinguish the third component of the pseudospin from that 
conventionally used for a bond along $c$ axis (${\mathcal T}_i^z$), we 
have labelled it with the $\zeta$ index. Note however  
that in contrast to the spin operator, the $\zeta$-component of the  
pseudospin operator, 
\begin{equation}
{\mathcal T}^{\zeta}=\sum_i{\mathcal T}_{i}^{\zeta}, 
\label{eq:Tzeta}
\end{equation}
does not commute with the Hamiltonian (\ref{eq:H_deg}), so the states 
with different values of this observable mix with each other. 
Nevertheless, we will use its eigenvalues together with the 
$z$-component $S^z$ of the total spin operator $\textbf{S}$ to specify 
multiparticle states in terms of which we write the Hamiltonian. It is 
straightforward to construct explicitly all 28 states; they are listed 
in \ref{ap:states}. 

In the high-spin subspaces $S^z=\pm1$, the Hamiltonian (\ref{eq:H_deg}) 
is decomposed into a zero $2\times2$ matrix $H_{0\sigma}$ involving
the two $|\Psi_{\alpha\sigma}\rangle$ states with $T^{\zeta}=\pm 1$, 
and two $2\times2$ matrices:
\begin{equation}
H_{1\sigma}= \Big(
\langle\Psi_{1\sigma}^{+}|,\langle\Psi_{2\sigma}^{-}|\Big) 
\left(\begin{array}{cc} 
                  0           &  t_{\zeta\zeta} - t_{\xi\xi} \\  
 t_{\zeta\zeta} - t_{\xi\xi}  &  U-3J_H
\end{array}\right)
\left(\begin{array}{c} 
|\Psi_{1\sigma}^{+}\rangle \\ 
|\Psi_{2\sigma}^{-}\rangle
\end{array}\right),
\label{eq:T1}
\end{equation}
and, 
\begin{equation}
H_{2\sigma}=
\Big(\langle\Psi_{1\sigma}^{-}|,\langle\Psi_{2\sigma}^{+}|\Big)
\left(\begin{array}{cc} 
                 0           &  t_{\zeta\zeta} + t_{\xi\xi} \\  
 t_{\zeta\zeta} + t_{\xi\xi} &  U-3J_H
\end{array}\right)
\left(\begin{array}{c} 
|\Psi_{1\sigma}^{-}\rangle \\ 
|\Psi_{2\sigma}^{+}\rangle
\end{array}\right).
\label{eq:T2}
\end{equation}
Now we are left with the $S^z=0$ subspace. In the $S=1$ sector one
recovers the same eigenenergies: two zeros corresponding to the above
$T^{\zeta}=\pm 1$ localized states with $T^{\zeta}=\pm 1$, and the 
eigenvalues following from the matrices $H_{1\sigma}$, and 
$H_{2\sigma}$. In the $S=0$ subspace, using the states with 
$T^{\zeta}=\pm 1$, the following Hamiltonian matrices are found in
addition:
\begin{equation}
H_3  = \Big(\langle\Phi_{1}^{-}|,\langle\Phi_{2}^{+}|,
\langle\Phi_{3}^{+}|,\langle\Phi_{4}^{-}|\Big)
\left(\!\begin{array}{cccc}  
  0              & 2t_{\zeta\zeta}   &  0          &  0              \\
 2t_{\zeta\zeta} &      U            & J_H         &  0              \\
  0              &     J_H           &  U          &  2t_{\xi\xi}    \\
  0              &      0            & 2t_{\xi\xi} &  0                 
\end{array}\right)
\left(\begin{array}{c} 
|\Phi_{1}^{-}\rangle \\ 
|\Phi_{2}^{+}\rangle \\
|\Phi_{3}^{+}\rangle \\ 
|\Phi_{4}^{-}\rangle \\ 
\end{array}\!\right),
\label{eq:S3}
\end{equation}
and,
\begin{equation}
H_4  = \Big(\langle\Phi_{2}^{-}|,\langle\Phi_{3}^{-}|\Big) 
       \left(\begin{array}{cc} 
   U & J_H  \\   
 J_H & U    
\end{array}\right)
\left(\begin{array}{c} 
|\Phi_{2}^{-}\rangle \\
|\Phi_{3}^{-}\rangle 
\end{array}\right).
\label{eq:S4}
\end{equation}
Finally, in the sector with $T^{\zeta}=0$, the Hamiltonian matrices read,
\begin{equation}
H_5  = \Big(\langle\Phi_{7}^{+}|,\langle\Phi_{8}^{-}|\Big) 
       \left(\begin{array}{cc} 
   U-J_H & t_{\zeta\zeta} + t_{\xi\xi}  \\   
 t_{\zeta\zeta} + t_{\xi\xi}& 0   
\end{array}\right)
\left(\begin{array}{c} 
|\Phi_{7}^{+}\rangle \\
|\Phi_{8}^{-}\rangle 
\end{array}\right),
\label{eq:S5b}
\end{equation}

\begin{equation}
H_6  = \Big(\langle\Phi_{7}^{-}|,\langle\Phi_{8}^{+}|\Big) 
       \left(\begin{array}{cc} 
   U-J_H & t_{\zeta\zeta} - t_{\xi\xi}  \\    
 t_{\zeta\zeta} - t_{\xi\xi}& 0   
\end{array}\right)
\left(\begin{array}{c} 
|\Phi_{7}^{-}\rangle \\
|\Phi_{8}^{+}\rangle 
\end{array}\right).
\label{eq:S6b}
\end{equation}

\subsection{Eigenenergies for $t_{2g}$ orbitals}
\label{subsec:t_2g}

Let us consider first a model with two active and equivalent $t_{2g}$ 
orbitals. 
The third orbital may be neglected in various contexts: 
($i$) either in a $d^1$ configuration when a crystal field raises the 
energy of the third orbital above the other two we consider, or 
($ii$) in a $d^3$ low spin configuration when a crystal field lowers 
the energy of the third orbital below the other two (the former one 
being filled), or 
($iii$) in a $d^5$ low spin configuration when a crystal field lowers 
the energy of the third orbital below the other two in which case one 
meets the problem of one hole in two orbitals. From the few layered 
perovskites that may correspond to these cases let us mention 
NdSrCrO$_4$ \cite{San81}. We note that the distortions in layered 
perovskites differ from the ones in cubic perokskites, where they are 
typically trigonal. For this model every matrix in equations
\eref{eq:S3}-\eref{eq:S6b} can be easily diagonalized and the 
corresponding eigenvalues are listed in the \ref{ap:t2g}. For clarity 
we have used the symmetry and classified them into triplet ($S=1$) and 
singlet ($S=0$) subspaces.

To get more insight into the competition between the tendencies towards 
the antiferromagnetic (AF) and FM ground states, we now discuss the 
lowest energy eigenstates in the strong coupling (large $U$) limit. As 
expected, the lowest energy, 
\begin{equation}
E_{T_{0}}^{t_{2g}}=
\case{1}{2}\Bigl(U-3J_H-\sqrt{(U-3J_H)^2+16t^2}\Bigr), 
\label{eq:Tt2g}
\end{equation}
obtained by the diagonalization of the matrix in equation \eref{eq:T2}, 
belongs to spin triplet coexisting with the pseudospin singlet. In the 
case of strong on-site interorbital repulsion \mbox{($U-3J_H\gg t$)}, 
the lowest high-spin energy \eref{eq:Tt2g} reads,
\begin{equation}
E_{T_{0}}^{t_{2g}}\simeq -\frac{4t^2}{U-3J_H}.
\label{eq:Tlow}
\end{equation}
Note however that finite $J_H$ could significantly reduce the value of 
the interorbital repulsion $U-3J_H$, so that it would no longer be much  
larger than $t$. As a consequence, significant corrections to the above 
result obtained to second order in $t/(U-3J_H)$ are expected in this 
case. Analogously, the lowermost low-spin energy results from the 
matrix in equation \eref{eq:S3}, which is degenerate with the one 
obtained from the matrix in equation \eref{eq:S5b}, 
\begin{equation}
E_{S_{0}}^{t_{2g}}= 
-\case{1}{2} \Bigl(U^{} - J_H -\sqrt{(U -J_H)^2 + 16t^2}\Bigr) 
\simeq -\frac{4t^2}{U-J_H},
\label{eq:Slow}
\end{equation}
in the strong coupling regime.

Comparison of equation (\ref{eq:Slow}) with the lowest high-spin energy 
given by equation (\ref{eq:Tlow}) allows one to draw interesting 
conclusions about conditions required for ferromagnetism. It is apparent 
that $E_{S_{0}}^{t_{2g}}$ and $E_{T_{0}}^{t_{2g}}$ are degenerate for 
$J_H=0$. However, even infinitesimally small $J_H>0$ lifts this 
degeneracy and might give rise to spin ferromagnetism combined with the 
alternating orbital (AO) pseudospin correlation. 

\Table 
{\label{tab:EDEz0t2g} 
Lowest eigenenergies of the model (\ref{eq:H_deg}) for $t_{2g}$ orbitals at 
$U=8t$, obtained for the representative values of
$J_H=0$, $U/8$ and $U/4$. The eigenstates $|n\rangle$ are specified by 
the total spin $S_n$ and the expectation value of the $\zeta$-component 
of the total pseudospin $\mathcal{T}^{\zeta}$ \eref{eq:Tzeta}. Triplet 
states with $S_n=1$ have three components $S^z_{n}=\pm 1,0$.  
}
\begin{tabular}{@{}*{11}{l}}
\br 
\centre{3}{$J_H=0$}      & &  
\centre{3}{$J_H=U/8$ }   & & 
\centre{3}{$J_H=U/4$ }   \\ 
\mr  
\centre{1}{$E_n/t$} & \centre{1}{$S_n$} & \centre{1}{$T_n^{\zeta}$} & &
\centre{1}{$E_n/t$} & \centre{1}{$S_n$} & \centre{1}{$T_n^{\zeta}$} & & 
\centre{1}{$E_n/t$} & \centre{1}{$S_n$} & \centre{1}{$T_n^{\zeta}$} \\ 
\mr 
\-0.4721 & 1 & \m0  & & \-0.7016 &   1    &    \m0          &    
 &\0\-1.2361 &    1   & \m0        \\
	  & 0 & $\pm1$  &  & \-0.5311 & 0 & \m0 & &\0\-0.6056 & 0 & \m0 \\
 	  &    0   &   \m0        & &           &   0    &    \m0        &  
  &       &    0   &   \m0          \\
\br
\end{tabular}
\endTable

As an illustrative example, we present in \tref{tab:EDEz0t2g} the lowest
eigenenergies $E_n$ obtained from the ED of the Hamiltonian 
(\ref{eq:H_deg}) for the $t_{2g}$ orbitals in the strong coupling regime 
$U=8t$ for a few values of $J_H$. We have specified them in terms of the 
total spin $S_n=0,1$. Another quantity used for the classification of 
states is the $\zeta$-component of the total pseudospin operator, see
equation \eref{eq:Tzeta}. However, even in this case it is not a good 
quantum number being modified by the 'pair-hopping' processes from one 
orbital to the other present in equation \eref{eq:H_int}. Indeed, in 
matrix \eref{eq:S3}, a sector consisting of $|\Phi_1^{-}\rangle$ and 
$|\Phi_2^{+}\rangle$ states, carrying the $T^{\zeta}=-1$ pseudospin 
flavour, is coupled to the one in terms of $|\Phi_3^{+}\rangle$ and
$|\Phi_4^{-}\rangle$ states, carrying the opposite $T^{\zeta}=1$ 
flavour. Similarly, sectors with different  pseudospin flavours are 
mixed in matrix \eref{eq:S4}. Thus, one has to determine the expectation 
values of $\mathcal{T}^{\zeta}$ by a direct evaluation using the 
eigenstates of the Hamiltonian.

The ground state of the $t_{2g}$ model with finite $J_H$ is 
a spin triplet accompanied by a pseudospin singlet (\textit{cf}. 
\tref{tab:EDEz0t2g}). Notice however that its energy $-1.2361t$ obtained 
for $J_H=U/4$ differs vastly from the value $E_{T_0}^{t_{2g}}=-2t$ 
estimated roughly from equation \eref{eq:Tlow}. The reason of this 
discrepancy is the failure of the second order perturbation theory 
controlled by $t^2/(U-3J_H)$, being here of order $\mathcal{O}(t)$. 
Therefore, it can be used only for qualitative arguments in this 
regime, while it works better for smaller $J_H=U/8$, yielding there 
$-0.8t$, a value much closer to the exact energy $-0.7016t$.

\subsection{Eigenenergies for $e_{g}$ orbitals}
\label{subsec:e_g}

Turning now to the $e_g$ orbital model and recollecting the immobile
$|\xi\rangle$ orbital flavour with $t_{\xi\xi}=0$, one immediately 
notices two sets of two identical $2\times2$ subspaces spanned by the
$T^{\zeta}=0$ states, described by the matrices in equations 
\eref{eq:S5b}-\eref{eq:S6b}, which readily yield two doubly degenerate 
singlets (\textit{cf}.~\ref{ap:eg}). Diagonalizing the submatrix 
\eref{eq:S3} yields the three eigenvalues given in \ref{ap:eg}. The first 
of them (\ref{eq:Seg1}) corresponds to the lowest energy low-spin state. 
In the strong coupling regime, i.e, for $(U-J_H)\gg t$, one finds 
\begin{equation}
E_{S_{0}}^{e_{g}}\simeq -\frac{4t^2U}{U^2-J_H^2}.
\label{eq:Seglow}
\end{equation}
It has to be compared with the lowest eigenvalue of the $S^z=1$ 
subspace, the lowest energy doubly degenerate spin triplet, 
\begin{equation}
E_{T_{0}}^{e_{g}}=\case{1}{2}\Bigl(U-3J_H-\sqrt{(U-3J_H)^2+4t^2}\Bigr), 
\label{eq:Teg}
\end{equation}
obtained by the diagonalization of the matrices \eref{eq:T1} and 
\eref{eq:T2}. It corresponds, as in the model with $t_{2g}$ orbitals, 
to the pseudospin singlet ($T_n=0$). In the strong coupling regime it 
yields the lowest high-spin energy,
\begin{equation}
E_{T_{0}}^{e_{g}}\simeq -\frac{t^2}{U-3J_H}.
\label{eq:Teglow}
\end{equation}
Note that the energy is much higher than that for the $t_{2g}$ model,
see equation \eref{eq:Tlow}, as only one electron is mobile.

\Table{
\label{tab:EDEz0eg} 
The same as in \tref{tab:EDEz0t2g} but for $e_{g}$ orbitals, and four
representative values of Hund's exchange: $J_H=0$, $U/8$, $U/4$, and 
$U/3$.  }
\hskip -1.1cm
\begin{tabular}{@{}*{12}{l}}
\br 
\centre{3}{$J_H=0$}      &  
\centre{3}{$J_H=U/8$ }   &
\centre{3}{$J_H=U/4$ }   &
\centre{3}{$J_H=U/3$ }  \\ 
\mr  
\centre{1}{$E_n/t$}     & \centre{1}{$S_n$} &  \centre{1}{$T_n^{\zeta}$}  & 
\centre{1}{$E_n/t$}     & \centre{1}{$S_n$} &  \centre{1}{$T_n^{\zeta}$}  & 
\centre{1}{$E_n/t$}     & \centre{1}{$S_n$} &  \centre{1}{$T_n^{\zeta}$}  & 
\centre{1}{$E_n/t$}     & \centre{1}{$S_n$} &  \centre{1}{$T_n^{\zeta}$} \\
\mr 
\-0.4721       &    0   &  \m\-1       &  \-0.4784 &   0    &  \m\-0.9985 &  
\0\-0.4983     &    0   & \m\-0.9935   &  \0\-1.0  &   1    &  \m0.0  \\
\-0.1231       &    1   &   \m0        &  \-0.1926 &   1    &  \m0.0      &
\0\-0.4142     &    1   &   \m0.0      &           &   1    &  \m0.0      \\
	       &    1   &   \m0        &           &   1    &  \m0.0      & 
               &    1   &   \m0.0      & \0\-0.5204 &   0   &  \m\-0.9877 \\
	       &    0   &   \m0        &  \-0.1401 &   0    &  \m0.0      &
\0\-0.1623     &    0   &   \m0.0      & \0\-0.1813 &   0    &  \m0.0     \\
      	       &    0   &   \m0        &           &   0    &  \m0.0      & 
               &    0   &   \m0.0      &           &   0    &  \m0.0   \\
\br
\end{tabular}
\endTable

Based on the lowest energy excitations \eref{eq:Seglow} and 
\eref{eq:Teglow} one can easily notice a striking difference between 
$t_{2g}$ and $e_{g}$ orbitals with respect to the ground state. Indeed, 
in contrast to the $t_{2g}$ model where even infinitesimally small 
$J_H>0$ lifts the degeneracy between the lowest energy singlet and 
triplet excitations and stabilizes the FM spin correlation, one can 
expect that the singlet state with the energy $E_{S_{0}}^{e_{g}}$ 
\eref{eq:Seg1} remains the $e_g$ ground state up to $J_H\sim U/4$. 

The corresponding exact eigenenergies of the Hamiltonian (\ref{eq:H_deg}) 
with $e_{g}$ orbitals obtained in the strong coupling regime $U=8t$ for a 
few values of $J_H$ are listed in \tref{tab:EDEz0eg}. The data show that 
increasing $J_H$ diminishes the energy difference between the lowest 
singlet and the first two triplet excited states. Remarkably, however,
even for unrealistically large $J_H=U/4$, the $e_g$ ground state is 
still a singlet with almost fully occupied mobile pseudospin 
$|\zeta\rangle$ orbitals (\textit{cf}. \tref{tab:EDEz0eg}) contradicting 
our predictions from the strong coupling regime. It follows from
the approximate low-spin state energy $E_{S_0}^{e_{g}}$ 
\eref{eq:Seglow} which overestimates the tendency towards ferromagnetism 
due to the performed expansion, and the transition from the singlet to 
the high-spin state occurs only for $J_H\simeq 0.27U$. Indeed, the 
ground state found for $J_H=U/3$ is a triplet, with the energy well 
below that of the singlet state (see \tref{tab:EDEz0eg}).
Therefore, although Hund's exchange coupling $J_H$ is a driving force 
of the FM spin correlations in both models, it is decisively more 
efficient in the case of $t_{2g}$ orbitals, where infinitesimal $J_H$
stabilizes the high-spin state as both pseudospins are mobile. 

Actually, one can rewrite the interaction term of the Hamiltonian 
$H_{\mathnormal int}$ \eref{eq:H_int} as a superposition of 
interactions in different channels, involving the density, magnetization, 
orbital polarization, magnetic orbital polarization, and on-site orbital 
flip term, respectively \cite{Fre05}. In each
of them the interaction strength is different: it is systematically
attractive (repulsive) in the magnetic (density) channels, while it
turns repulsive in the orbital polarization channel for $5J_H>U$. As a
result, one expects a transition in the ground state from pseudospin 
triplet to singlet with increasing $J_H/U$. This is precisely the 
transition we discussed above for $e_g$ electrons. In contrast,
the magnetic instability takes over for $t_{2g}$ electrons, 
and the spin triplet ground state is found at any $J_H>0$.

\subsection{Correlation functions and susceptibility at orbital degeneracy}

We now turn to the temperature dependence of the on-site 
$\langle{\textbf S}_i^2\rangle$ and intersite  
$\langle{\textbf S}_1\cdot{\textbf S}_{2}\rangle$ spin correlations,
and the on-site $\langle{\textbf T}_i^2\rangle$ and intersite 
$\langle{\textbf T}_1\cdot{\textbf T}_{2}\rangle$ pseudospin 
correlations. The latter yield information about an orbital state 
together with orbital correlation between neighbouring sites. 

\begin{figure}[t!]
\begin{center}
\includegraphics*[width=0.72\textwidth ]{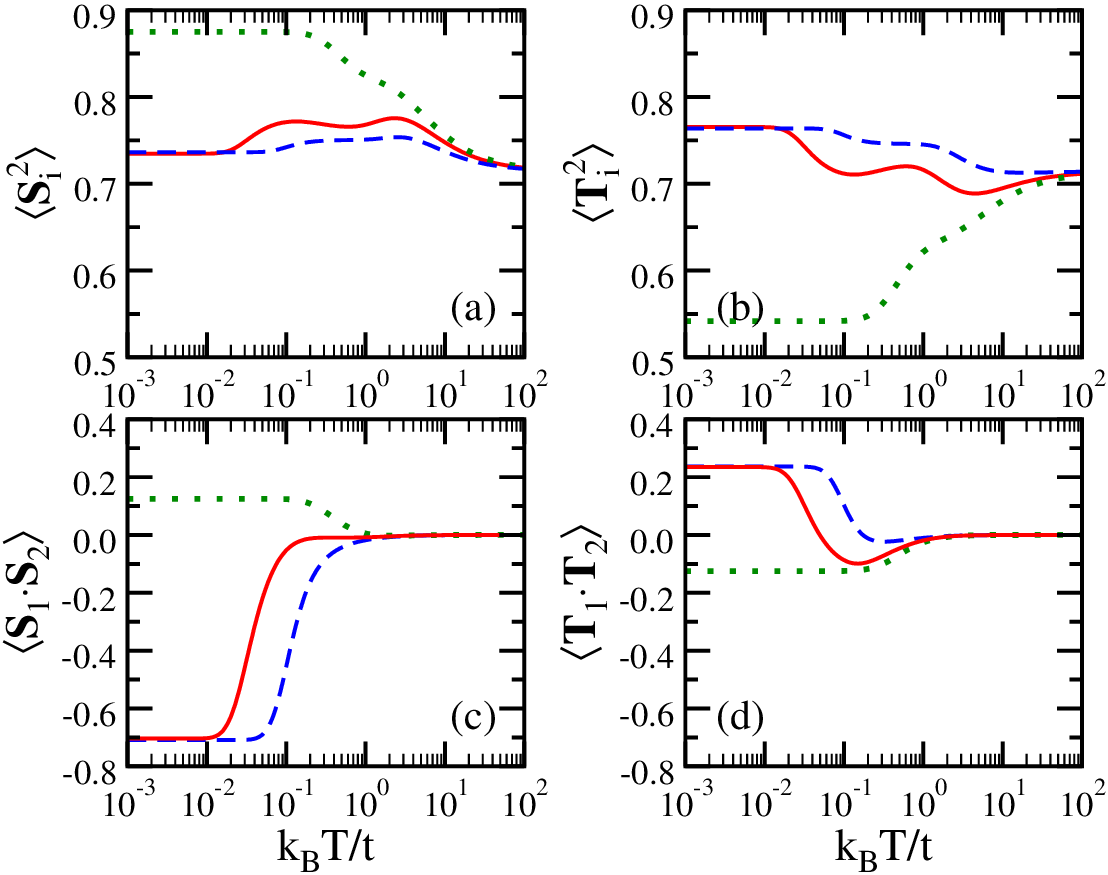}
\end{center}
\caption{ 
Temperature dependence of local moments 
(a) spin $\langle{\textbf S}_i^2\rangle$, 
(b) pseudospin $\langle{\textbf T}_i^2\rangle$, 
and of intersite correlations
(c) spin $\langle{\textbf S}_1\cdot{\textbf S}_{2}\rangle$,
(d) pseudospin $\langle{\textbf T}_1\cdot{\textbf T}_{2}\rangle$, 
as obtained for the model with $e_g$ orbitals, for increasing values of 
Hund's exchange: $J_H=U/8$ (dashed line), $J_H=U/4$ (solid line) and
$J_H=U/3$ (dotted line). 
Parameters: $U=8t$, $E_0=0$. }
\label{fig:Ez0eg}
\end{figure}

In \fref{fig:Ez0eg} we present the temperature dependence of the spin 
and pseudospin correlation functions as well as both susceptibilities of 
the model with $e_g$ orbitals. We have set Hund's exchange coupling to 
be $J_H/U=1/8$ (dashed line) and $J_H/U=1/4$ (solid line).
At low temperature, one expects charge localization as the system is in 
the strong coupling regime $U=8t$. Indeed, the local spin moment 
$\langle{\textbf S}_i^2\rangle$ reaches virtually the magnitude 
$S(S+1)=3/4$ for the spin $S=1/2$. A rise of 
$\langle{\textbf S}_i^2\rangle$ above this value upon increasing 
temperature is caused by thermal excitations to triplet states. 
They are favored by Hund's interaction and form local high-spin 
configurations. Consequently, the rise of $\langle{\bf S}_i^2\rangle$ is 
larger for stronger $J_H=U/4$. Next, the intersite spin correlation 
function $\langle{\textbf S}_1\cdot{\textbf S}_2\rangle$ indicates the 
low-spin (AF) nature of the ground state, whereas the corresponding 
pseudospin function $\langle{\textbf T}_1\cdot{\textbf T}_{2}\rangle$ 
illustrates the ferro orbital (FO) pseudospin correlation. 

\begin{figure}[t!]
\begin{center}
\includegraphics*[width=0.72\textwidth ]{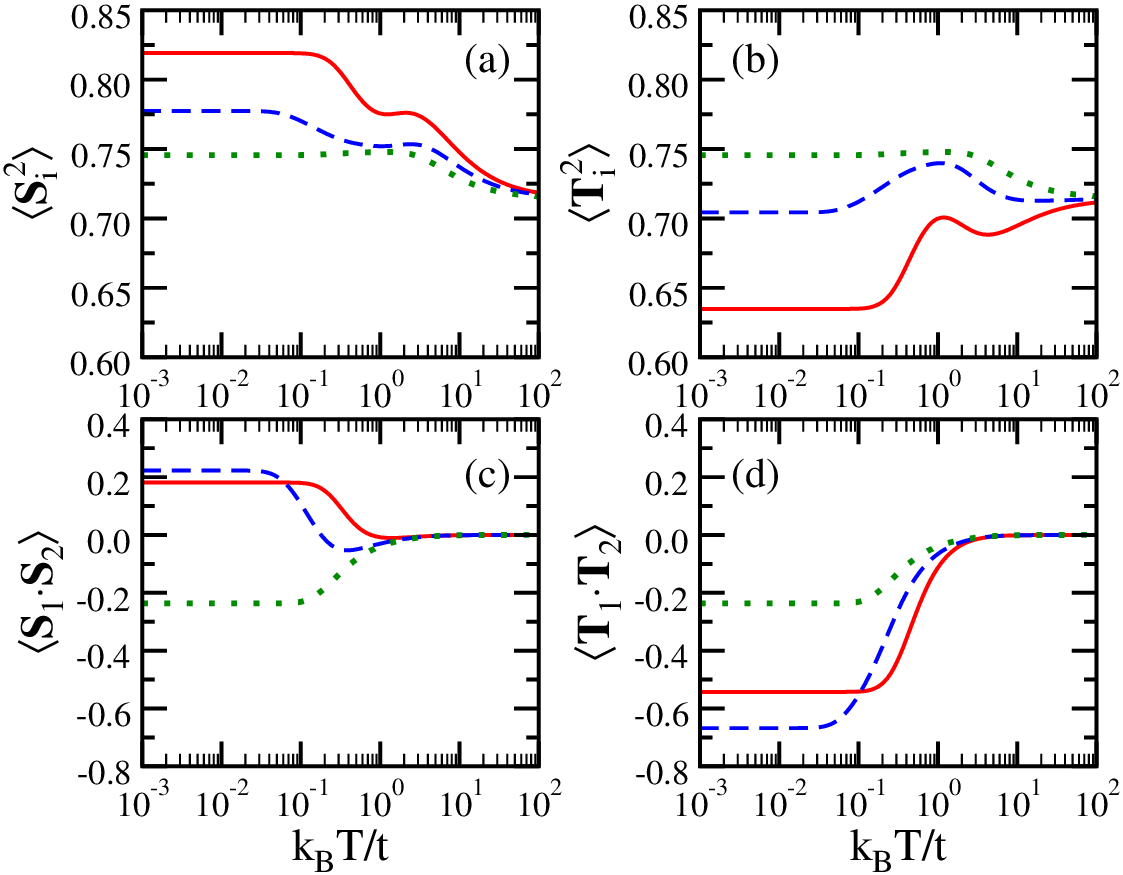}
\end{center}
\caption{ 
The same as in \fref{fig:Ez0eg} but for the model with $t_{2g}$ 
orbitals, and for: 
$J_H=0$ (dotted line), $J_H=U/8$ (dashed line), $J_H=U/4$ (solid line).
}
\label{fig:Ez0t2g}
\end{figure}

The results obtained within the $t_{2g}$ model are qualitatively 
different. First, at $J_H=0$, apart from spin SU(2) symmetry there is an 
additional SU(2) symmetry for orbital degrees of freedom, resulting in
higher SU(4) symmetry \cite{su4} and intersite spin and pseudospin 
correlation functions are both negative and identical. Second, at 
finite $J_H$ positive
$\langle{\textbf S}_1\cdot{\textbf S}_{2}\rangle$ indicates the FM 
nature of the ground state supported by the pseudospin singlet with  
negative $\langle{\textbf T}_1\cdot{\textbf T}_{2}\rangle$ correlations 
shown in \fref{fig:Ez0t2g}. 

\begin{figure}[b!]
\begin{center}
\includegraphics*[width=0.72\textwidth ]{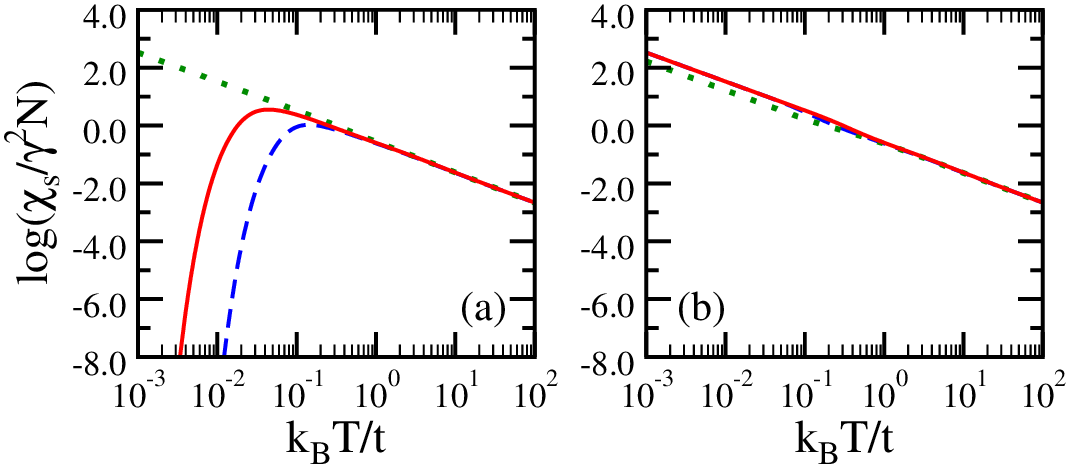}
\end{center}
\caption{ 
Temperature dependence of the spin susceptibility 
$\log(\chi_s/\gamma^2N)$ per site ($N=2$) as obtained for two values of 
Hund's exchange, $J_H=U/8$ (dashed lines) and $J_H=U/4$ (solid lines), 
for: 
(a) $e_g$ model, and 
(b) $t_{2g}$ model.
For comparison the susceptibilities obtained with $J_H=U/3$ for the 
$e_g$ model and with $J_H=0$ for the $t_{2g}$ model are shown by dotted 
lines. Parameters: $U=8t$, $E_0=0$. }
\label{fig:chiEz0}
\end{figure}

The gradual increase of triplet intersite correlations  
$\langle{\textbf S}_1\cdot{\textbf S}_{2}\rangle$ observed for the 
$e_g$ model is also well recognized in the spin susceptibility, see 
\fref{fig:chiEz0}(a). Upon taking the logarithm of $\chi_s$ we find a 
typical AF behaviour with a characteristic cusp at the crossover 
temperature $T_c(e_g)$. Obviously, the AF array that sets in has zero 
net magnetic moment at the temperature below $T_c(e_g)$ and this 
explains the observed cusp in $\chi_s$. Only for large $J_H=U/3$ the 
ground state is a triplet, and the spin susceptibility changes to the 
Curie-Weiss type. In contrast, in case of $t_{2g}$ model 
[\fref{fig:chiEz0}(b)], $\chi_s$ is large at low temperature already for 
$J_H=0$ as the ground state has degenerate singlet and triplet states. 
Increasing $J_H$ gives here only moderate enhancement of the spin 
susceptibility as the degeneracy of the ground state is then removed. 

\subsection{Specific heat and entropy at orbital degeneracy}

\begin{figure}[t!]
\begin{center}
\includegraphics*[width=0.75\textwidth ]{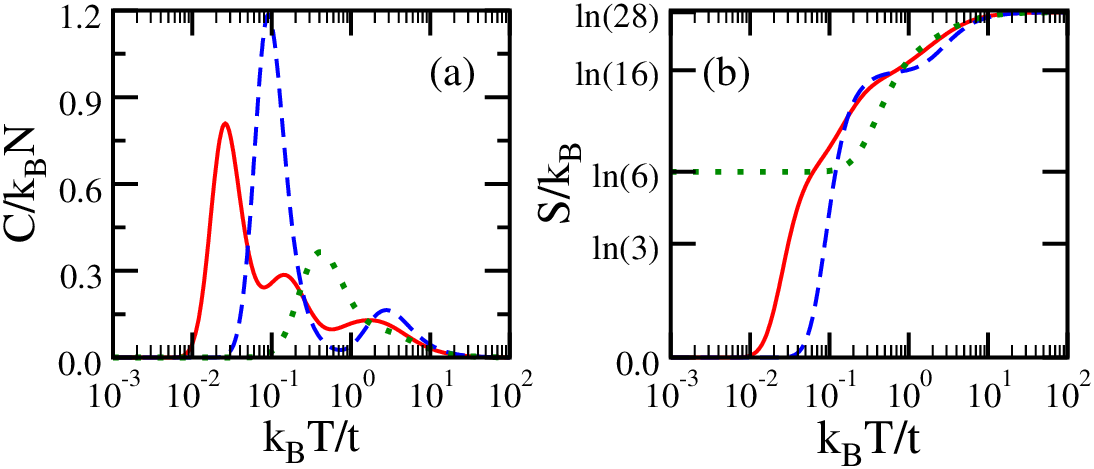}
\end{center}
\caption
{ Temperature dependence of 
(a) the specific heat $C$ per site and 
(b) the entropy $S$, as obtained for $e_g$ orbitals with:
$J_H=U/8$ (dashed line), $J_H=U/4$ (solid line), and $J_H=U/3$ 
(dotted line). Parameters: $U=8t$, $E_0=0$.
}
\label{fig:SEz0eg}
\end{figure}

\begin{figure}[t!]
\begin{center}
\includegraphics*[width=0.75\textwidth ]{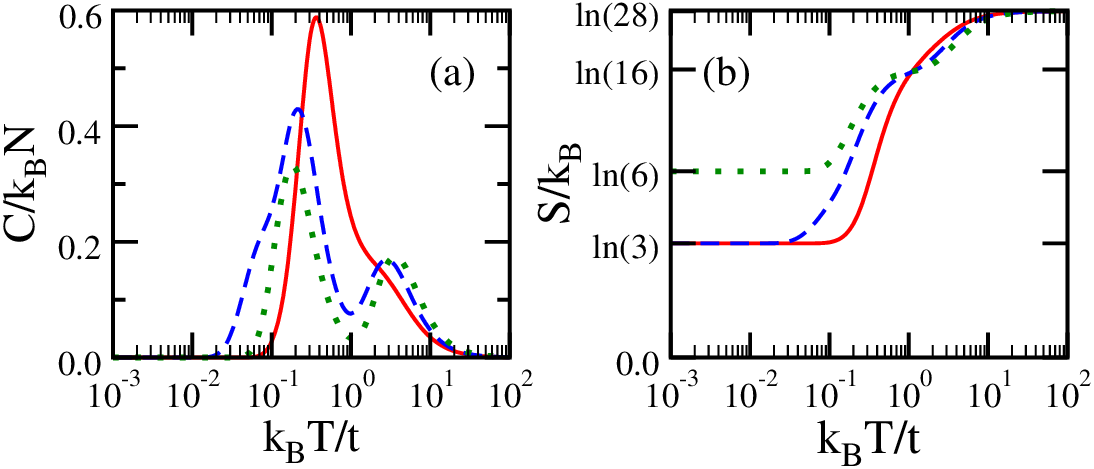}
\end{center}
\caption
{ 
The same as in \fref{fig:SEz0eg} but for $t_{2g}$ orbitals. 
Dotted, dashed, and solid line corresponds to $J_H=0$, $J_H=U/8$, and 
$J_H=U/4$, respectively. }
\label{fig:SEz0t2g}
\end{figure}

Different energy spectra of the $e_g$ and $t_{2g}$ systems result in 
quite different temperature behaviour of the specific heat $C$ and the 
entropy $S$, as shown in figures~\ref{fig:SEz0eg} and \ref{fig:SEz0t2g}.
Consider first the $e_g$ system. As depicted in \fref{fig:SEz0eg}, a low 
temperature peak of the specific heat coincides with the characteristic 
kink in the susceptibility $\chi_s$ [\textit{cf}. \fref{fig:chiEz0}(a)]. 
Comparing the position of the peaks corresponding to $J_H=U/8$ and 
$J_H=U/4$, one finds that increasing $J_H$ reduces $T_c(e_g)$, as the 
crossover to the high-spin state is approached. Note, however, that for 
$J_H=U/4$ the low temperature peak in the specific heat splits into two. 
The first one corresponds to a transition from the singlet ground state 
to the first two triplet excited states with the excitation energy 
$\Delta E_1/t=0.0841$, whereas the second one appears due to the higher 
energy charge excitations, with the excitation energies 
$\Delta E_2/t=0.336$ and $\Delta E_3/t=0.498$, respectively. 
In contrast, when $J_H=U/8$ the excitation energy into the first excited 
state is much larger $\Delta E_1'/t=0.286$, whereas the other excitation 
energies are nearly unaltered: $\Delta E_2'/t=0.338$ and 
$\Delta E_3'/t=0.478$. This results in a single broad low temperature 
peak. High temperature peaks occur due to thermal excitations which 
create double occupancies and lead to charge delocalization, well seen 
in the suppression of $\langle{\textbf S}_i^2\rangle$. Finally, at large
$J_H=U/3$ all peaks in the specific heat merge into a single broad peak 
at higher temperature.

Turning now to the temperature dependence of the entropy depicted in 
\fref{fig:SEz0eg}(b), one finds $S\simeq 0$ in the low $T$ regime for 
the singlet ground state. Basically, the overall rapid increase of the 
entropy around $T_c(e_g)$ in \fref{fig:SEz0eg}(b) is very much the same 
for $J_H=U/8$ and $J_H=U/4$, corresponding to the low temperature peak 
in the specific heat. However, a detailed behaviour of $S$ depends on 
$J_H$. In the regime of small Hund's exchange $J_H\le U/8$, where 
$L_S=16$ singly occupied states are well separated from doubly occupied 
ones, $S$ possesses a point of inflection $S=k_B\ln 16$ at 
$k_BT\simeq t$, which follows from the gapped character of the specific 
heat. In contrast, for the larger $J_H=U/4$, the entropy increase starts 
at lower temperature as the energy of the singlet-triplet excitation is 
low. Here the gap between singly and doubly occupied states is smaller 
and the corresponding point of inflection is less transparent. The 
limiting value $S=k_B\ln 28$ results from the calculation performed in 
the canonical ensemble.

The situation is quite different in the case of the $t_{2g}$ model. 
Unlike the $e_g$ case, increasing $J_H$ shifts $T_c(t_{2g})$ towards 
higher temperatures (\fref{fig:SEz0t2g}). As a result, the crossover 
temperatures of both systems differ substantially, especially in the 
large $J_H=U/4$ regime. Indeed, from the position of the low temperature 
peak of the specific heat in figures~\ref{fig:SEz0eg}(a) and 
\ref{fig:SEz0t2g}(a), one can read off that $k_BT_c(e_{g})=0.025t$, 
whereas $k_BT_c(t_{2g})=0.35t$. The origin of this marked difference is 
certainly the fact that increasing $J_H$ diminishes (enlarges) the gap 
between the spin singlet (triplet) ground state and the first excited 
triplet (singlet) state of the $e_g$ ($t_{2g}$) system, respectively 
(\textit{cf}. tables~\ref{tab:EDEz0t2g} and \ref{tab:EDEz0eg}). 
In contrast to the $e_g$ case, the entropy is finite in the low 
temperature regime, and approaches either the value $k_B\ln 3$ for both 
$J_H=U/8$ and $J_H=U/4$ or $k_B\ln 6$ when $J_H=0$, see 
\fref{fig:SEz0t2g}(b). Indeed, for $J_H=0$, the ground state corresponds 
to six degenerate states --- three of them constitute a spin triplet, 
whereas the others are singlets, as shown in \tref{tab:EDEz0t2g}.
However, any finite $J_H>0$ splits up these states and leads to the 
triplet ground state with the entropy $S=k_B\ln 3$. As in the $e_g$ 
case, $S$ has a clear point of inflection only in the small $J_H\le U/8$ 
regime.

\begin{figure}[t!]
\begin{center}
\includegraphics*[width=0.75\textwidth ]{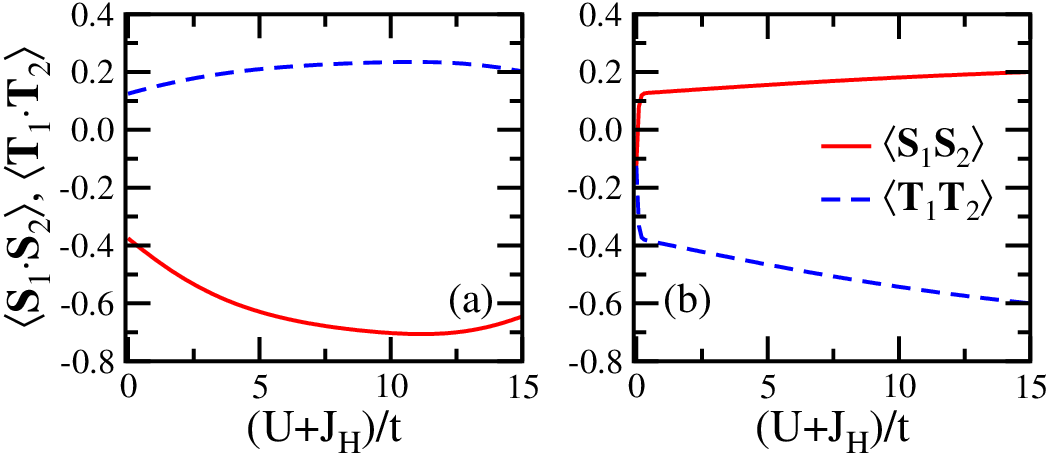}
\end{center}
\caption{ 
Zero temperature intersite correlation functions: 
spin $\langle{\textbf S}_1\cdot{\textbf S}_{2}\rangle$ (solid lines) 
and pseudospin $\langle{\textbf T}_1\cdot{\textbf T}_{2}\rangle$ 
(dashed lines) as functions of the Stoner parameter $U+J_H$ 
in the two-band model at $J_H=U/4$ for: 
(a) $e_g$ orbitals, and 
(b) $t_{2g}$ orbitals. 
}
\label{fig:fUJHEz0}
\end{figure}

To summarize this section, we compare in \fref{fig:fUJHEz0} the 
intersite spin $\langle{\textbf S}_1\cdot{\textbf S}_{2}\rangle$ and 
pseudospin $\langle{\textbf T}_1\cdot{\textbf T}_{2}\rangle$ 
correlation functions for the $e_g$ system [\textit{cf}. 
\fref{fig:fUJHEz0}(a)] and for the $t_{2g}$ system [\textit{cf}. 
\fref{fig:fUJHEz0}(b)] as a functions of the Stoner parameter $U+J_H$. 
Note that $\langle{\textbf S}_1\cdot{\textbf S}_{2}\rangle$ is finite 
and negative even in the noninteracting $U=0$ limit due to the Pauli 
principle. On the one hand, the $e_g$ results illustrate the AF  
correlations in the ground state (negative 
$\langle{\textbf S}_1\cdot{\textbf S}_{2}\rangle$) owing to the 
preferred mobile pseudospin $|\zeta\rangle$. On the other hand, the 
ground state of the $t_{2g}$ system with a finite interaction is a spin 
triplet as $\langle{\textbf S}_1\cdot{\textbf S}_{2}\rangle$ is positive 
and a pseudospin singlet as the corresponding correlation function 
$\langle{\textbf T}_1\cdot{\textbf T}_{2}\rangle$ is negative. 
We therefore conclude that the ground state properties strongly depend 
on the orbital correlations, in agreement with the Goodenough-Kanamori
rules \cite{Goode}.

\section{New features at finite crystal field}
\label{sec:nondeg}

\subsection{Correlation functions and susceptibility at finite crystal 
field}
 
We now discuss the influence of the crystal field splitting given by 
equation (\ref{eq:H_cf}). Here we are interested in the nontrivial case 
of crystal field acting along the $c$ axis \emph{perpendicular} to the 
chain. Hence, as we have been working with the $e_g$ orbital basis 
consisting of a directional orbital along the molecular bond 
$|\zeta\rangle$ and an orthogonal to it planar orbital $|\xi\rangle$, 
one needs to rotate the field (\ref{eq:H_cf}) expressed in the original 
orbital basis $\{|x\rangle,|z\rangle\}$ by the same angle 
$\theta=2\pi/3$ which enabled us to simplify the form of the hopping 
matrix \eref{eq:H_kin} into \eref{eq:H_eg}. Making the inverse
transformation to \eref{eq:rot} in equation \eref{eq:H_cf}, one obtains 
the crystal field term,
\begin{equation}
H_{\mathnormal cf} = \case{1}{2}E_0\sum_{i\sigma}\Bigl[
\cos\theta(c_{i\xi\sigma}^{\dag}c_{i\xi\sigma}^{}
-c_{i\zeta\sigma}^{\dag}c_{i\zeta\sigma}^{})
+\sin\theta(c_{i\xi\sigma}^{\dag}c_{i\zeta\sigma}^{}
+c_{i\zeta\sigma}^{\dag}c_{i\xi\sigma}^{})\Bigr].
\label{eq:H_cfrot}
\end{equation}
Note that $\cos\theta$ is negative so the field \eref{eq:H_cfrot}  
favours the $|\xi\rangle$ occupancy over the $|\zeta\rangle$ occupancy,  
as it should.

\Table{
\label{tab:EDEz2t2g} 
Eigenenergies of model (\ref{eq:H_deg}) with $t_{2g}$ orbitals as
obtained at $U=8t$ with a finite crystal 
field $E_{0}=2t$ acting perpendicular to the chain, for two values of
Hund's exchange: $J_H=U/8$ and $J_H=U/4$. The spin quantum number $S_n$
and the expectation value of $T_n^{\zeta}$ operator \eref{eq:Tzeta} in
each eigenstate $|n\rangle$ are given. }
\begin{tabular}{@{}*{7}{l}}
\br 
\centre{3}{$J_H=U/8$ }   & &  
\centre{3}{$J_H=U/4$ }   \\ 
\mr
\centre{1}{$E_n/t$} & \centre{1}{$S_n$} & \centre{1}{$T_n^{\zeta}$}& & 
\centre{1}{$E_n/t$} & \centre{1}{$S_n$} & \centre{1}{$T_n^{\zeta}$}\\
\mr 
\0\-2.4767 &    0   &  0.4995   & & \0\-2.4910  &   0    &    0.4980 \\
\0\-2.0    &    1   &  0.5      & & \0\-2.0     &   1    &    0.5    \\
\0\-0.7016 &    1   &  0.0      & &\0\-1.2361   &   1    &    0.0    \\
\0\-0.5311 &    0   &  0.0      & &\0\-0.6056   &   0    &    0.0    \\
\00.0      &    1   &  0.0      & & \00.0       &   1    &    0.0    \\
           &    0   &  0.0      & &             &   0    &    0.0    \\
\01.5183   &    0   &  \-0.4977 & & \01.4876    &   0    &   \-0.4899 \\
\02.0      &    1   &  \-0.5    & & \02.0       &   1    &   \-0.5   \\
\05.0      &   1    &   0.0     & &             &   1    &    0.0    \\
\05.7016   &   1    &   0.0     & & \03.2361    &   1    &    0.0    \\
\05.7639   &   0    &  0.4472   & & \05.1716    &   0    &   0.3536  \\
\06.2695   &   0    &  0.4498   & & \05.7610    &   0    &   0.3539  \\
\07.0      &   0    &   0.0     & &  \06.0      &   0    &   0.0     \\
\07.5311   &   0    &   0.0     & & \06.6056    &   0    &   0.0     \\
10.2361  &   0    & \-0.4472    & & 10.8284     &   0    & \-0.3536  \\
10.6890  &   0    & \-0.4516    & & 11.2424     &   0    &  \-0.3620 \\ 
\br
\end{tabular}
\endTable

Unfortunately, except for the Hamiltonian matrix \eref{eq:T2}, the form 
of the other matrices is now considerably more involved due to 
offdiagonal elements in the crystal field in equation \eref{eq:H_cfrot}
which couple states with different $T^{\zeta}$. 
In general, it is not possible to obtain analytic expressions for 
the eigenvalues and one has to resort to a numerical diagonalization. 
However, due to the equivalent hopping amplitudes, the eigenvalues of 
the $t_{2g}$ model should be independent of the rotation angle $\theta$, 
i.e., one has to get the same energy spectrum for the field 
\eref{eq:H_cfrot} with finite $\theta$, as well as for the diagonal in 
pseudospin space field of the form $E_0\mathcal{T}^{\zeta}$. It acts 
along the chain and corresponds to $\theta=0$ in equation 
\eref{eq:H_cfrot}.  

\Table {\label{tab:EDEz2eg} 
The same as in \tref{tab:EDEz2t2g} but for $e_{g}$ orbitals.  }
\begin{tabular}{@{}*{7}{l}}
\br 
\centre{3}{$J_H=U/8$ }   &  & 
\centre{3}{$J_H=U/4$ }   \\ 
\mr
\centre{1}{$E_n/t$} & \centre{1}{$S_n$} &  \centre{1}{$T_n^{\zeta}$} & & 
\centre{1}{$E_n/t$} & \centre{1}{$S_n$} &  \centre{1}{$T_n^{\zeta}$}\\
\mr 
\0\-2.0786  &   0    &  0.4229    & &  \0\-2.0965  &   1    &  0.4429 \\
\0\-2.0548  &   1    &  0.4695    & &  \0\-2.0852  &   0  &    0.4179 \\
\0\-0.2286  &   0    & \-0.0348   & &  \0\-0.4142  &   1  &    0.0    \\
\0\-0.1926  &   1    &   0.0      & &  \0\-0.2410  &   0  & \-0.0305  \\
\0\-0.1468  &   0    &    0.0002  & &  \0\-0.1703  &   0  &    0.0003 \\
\0\-0.0494  &   1    &    0.0734  & &  \0\-0.1169  &   1  &    0.1658 \\
\01.6727    &   0    &  \-0.3841  & &  \01.5089    &   1  &  \-0.4969 \\
\01.8878    &   1    & \-0.5409   & &  \01.6340    &   0  &  \-0.3713 \\
\05.1926    &   1    &   0.0      & &  \02.4142    &   1  &   0.0     \\
\05.2165    &   1    & \-0.0020   & &  \02.7045    &   1  &  \-0.1118 \\
\05.8526    &   0    & 0.4929     & &  \05.2848    &   0  &  0.4140   \\
\05.8564    &   0    &   0.4885   & &  \05.3156    &   0  &  0.3627   \\
\07.0374    &   0    & \-0.0538   & &  \06.0468    &   0  &  \-0.0631 \\
\07.2305    &   0    &    0.0461  & &  \06.2702    &   0  &   0.0517  \\
10.2567     &   0    &  \-0.4394  & &  10.8387     &   0  &  \-0.3512 \\
%
10.5476     &   0    &  \-0.5385  & &  11.1064     &   0  &  \-0.4305 \\ 
\br
\end{tabular}
\endTable

Numerical values of the eigenvalues obtained for $t_{2g}$ orbitals 
for $E_0=2t$ and $\theta=2\pi/3$ are listed in \tref{tab:EDEz2t2g}. 
Although the energy spectrum of the $t_{2g}$ model indeed does not 
depend on the field direction, expectation values of 
$\mathcal{T}^{\zeta}$ in the Hamiltonian eigenstates certainly do, 
as a finite rotation angle $\theta$ enables the mixing of states with 
different values of $\mathcal{T}^{\zeta}$. For example, the initial 
pseudospin $T^{\zeta}=\pm 1$ of spin triplets with the energy 
$E_0/t=\pm 2$ is reduced up to $T^{\zeta}=\pm 0.5$ by the field with 
$\theta=2\pi/3$ (\textit{cf}. \tref{tab:EDEz2t2g}), whereas it is 
conserved when $\theta=0$.

Contrary to the $E_0=0$ case with a spin triplet and pseudospin singlet 
as the $t_{2g}$ ground state at finite $J_H>0$, finite positive $E_0$  
suppresses the AO pseudospin correlation and stabilizes a spin singlet 
with a positive value of $T^{\zeta}\simeq 0.5$, inducing FO 
correlations. Remarkably, the effect of the crystal field on the $e_g$ 
ground state is just the opposite, as reported in \tref{tab:EDEz2eg}. 
Namely, by lifting the 
degeneracy of pseudospin flavours it promotes  the immobile 
$|\xi\rangle$ one. Consequently, there is not that much kinetic energy 
to be gained and the Coulomb interactions start to be crucial. However, 
they are noticeably better optimized by the FM spin correlation. 
Indeed, from \tref{tab:EDEz2eg} one sees a strong competition between  
the lowest singlet and triplet states, both with a positive but still 
smaller than in the $t_{2g}$ case value of $T^{\zeta}\simeq 0.45$. 
Note also that it becomes energetically advantageous to have the spin 
triplet as the ground state for large $J_H=U/4$ while a smaller Hund's 
exchange coupling $J_H=U/8$ drives the system towards the singlet in the 
ground state.   

\begin{figure}[t!]
\begin{center}
\includegraphics*[width=0.72\textwidth ]{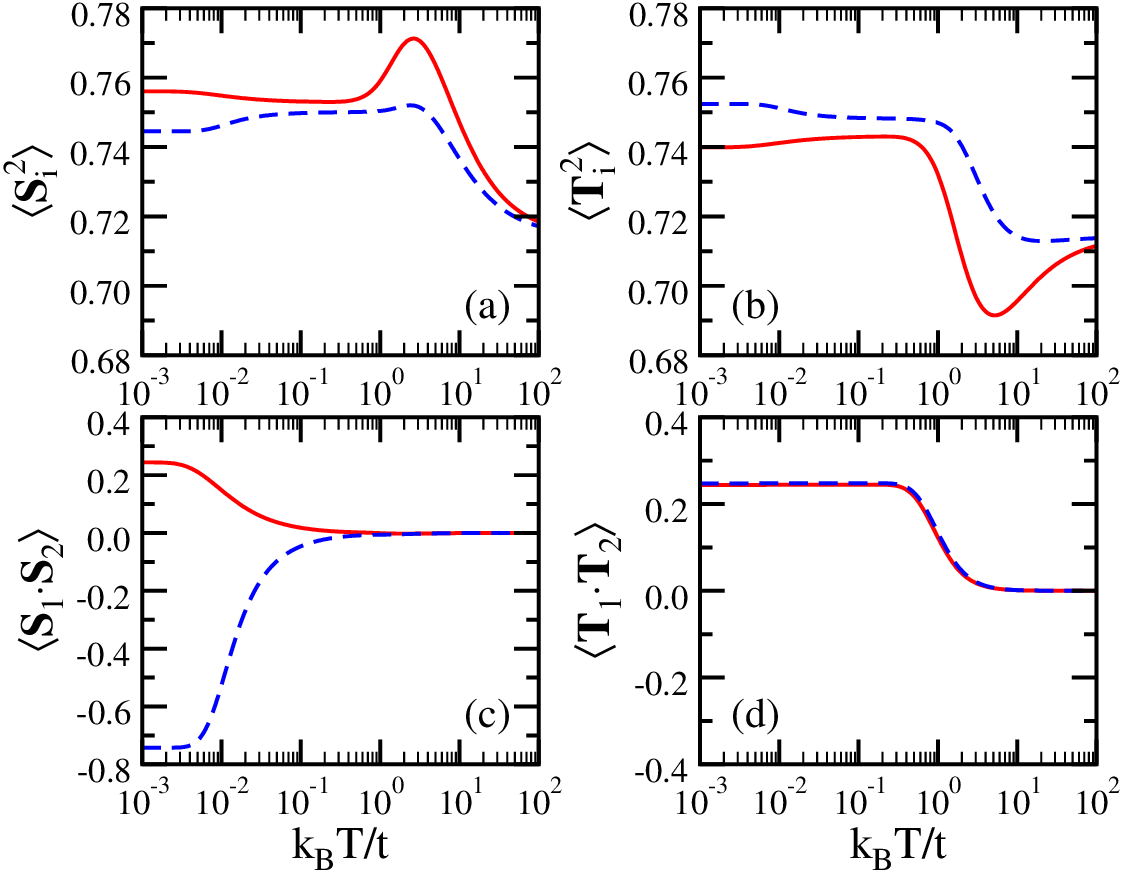}
\end{center}
\caption{ 
Temperature dependence of  local moments 
(a) spin $\langle{\textbf S}_i^2\rangle$ and 
(b) pseudospin $\langle{\textbf T}_i^2\rangle$, and of intersite
correlation functions
(c) spin $\langle{\textbf S}_1\cdot{\textbf S}_{2}\rangle$ and 
(d) pseudospin $\langle{\textbf T}_1\cdot{\textbf T}_{2}\rangle$, 
as obtained for the model with $e_g$ orbitals for: 
$J_H=U/8$ (dashed lines) and $J_H=U/4$ (solid lines). 
Parameters: $U=8t$ and $E_0=2t$. }
\label{fig:Ez2eg}
\end{figure}

The above ground states and the excitation spectra obtained at finite 
crystal field $E_0=2t$ influence intersite spin and orbital 
correlations, shown in \fref{fig:Ez2eg}. The ground state of the $e_g$ 
system depends on the value of $J_H$, being a spin singlet for $J_H=U/8$ 
which results in negative 
$\langle{\textbf S}_1\cdot{\textbf S}_2\rangle$ and a spin triplet for 
$J_H=U/4$ yielding positive 
$\langle{\textbf S}_1\cdot{\textbf S}_2\rangle$ in the low temperature
regime. In contrast, $\langle{\textbf T}_1\cdot{\textbf T}_2\rangle$ is 
positive (FO pseudospin correlations) and almost 
insensitive to the value of $J_H$ at low temperature. 
For increasing temperature one finds two transitions --- firstly, the 
intersite spin correlations weaken, and secondly pseudospin correlations 
weaken at much higher temperature, when charge excitations are 
thermally activated. This separation of the energy scales for spin and
orbital excitations occurs both for lower $J_H=U/8$ and higher 
$J_H=U/4$. 

\begin{figure}[b!]
\begin{center}
\includegraphics*[width=0.72\textwidth ]{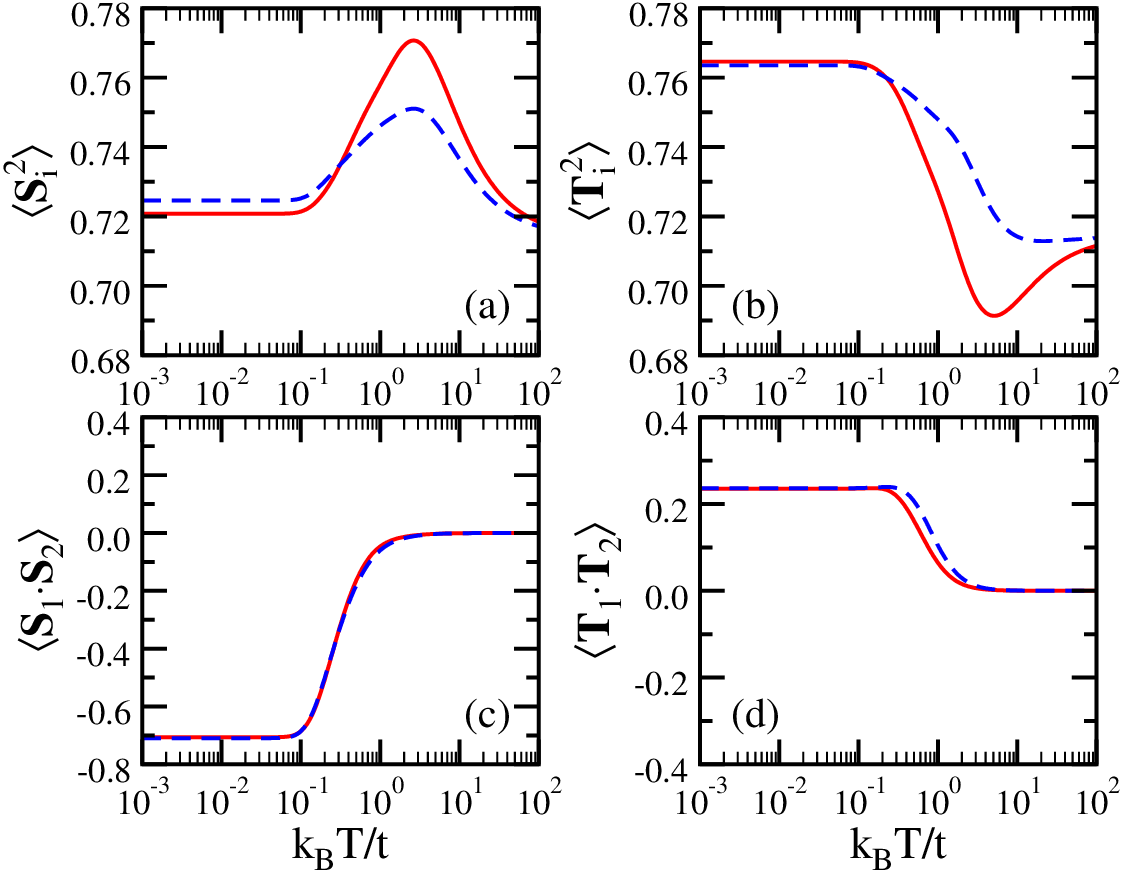}
\end{center}
\caption{ 
The same as in \fref{fig:Ez2eg} but for the model with $t_{2g}$ 
orbitals.
Dashed (solid) line corresponds to $J_H=U/8$ ($J_H=U/4$), respectively. 
}
\label{fig:Ez2t2g}
\end{figure}

A finite crystal field affects drastically the behaviour of the 
$t_{2g}$ correlation functions as well (\fref{fig:Ez2t2g}). At low 
temperature negative $\langle{\bf S}_1\cdot{\bf S}_2\rangle$ reveals 
the AF coupling between spins, whereas positive 
$\langle{\textbf T}_1\cdot{\textbf T}_2\rangle$ indicates the FO 
pseudospin correlation, regardless of $J_H$. In contrast to the $e_g$
case, however, spin and pseudospin correlations are reduced and vanish 
simultaneously, indicating that spin-orbital degrees of freedom are
stronger coupled in this case \cite{Kha00}.

\begin{figure}[t!]
\begin{center}
\includegraphics*[width=0.72\textwidth ]{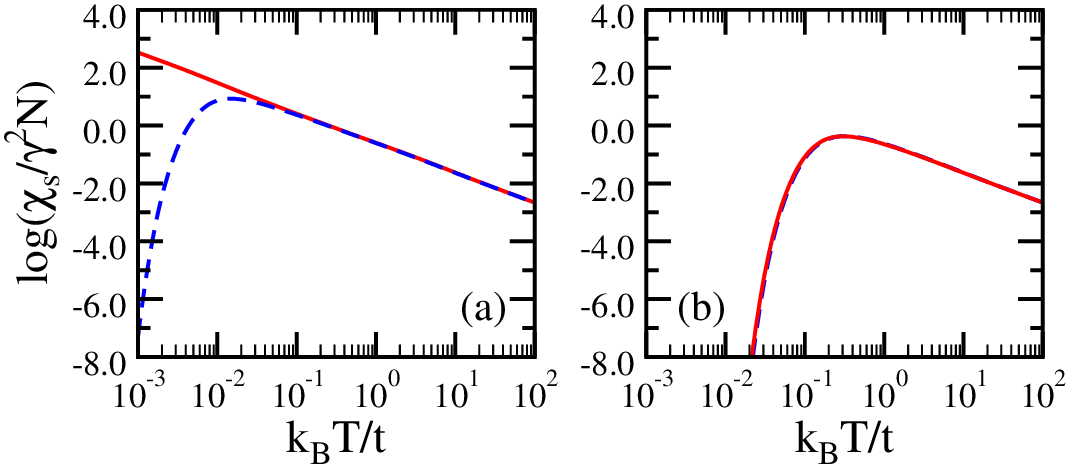}
\end{center}
\caption{ 
Temperature dependence of the spin susceptibility 
$\log(\chi_s/\gamma^2N)$ per site ($N=2$) for: 
(a) $e_g$ model, and 
(b) $t_{2g}$ model,
as obtained for two values of Hund's exchange: 
$J_H=U/8$ (dashed lines) and $J_H=U/4$ (solid lines).
Parameters: $U=8t$, $E_0=2t$. }
\label{fig:chiEz2}
\end{figure}

The change of the magnetic correlations with increasing $J_H$ in the 
$e_g$ model is reflected in the crossover from the AF behaviour of 
$\chi_s$ at $J_H=U/8$ to the Curie-Weiss behaviour of $\chi_s$ obtained 
for $J_H=U/4$, see \fref{fig:chiEz2}(a). In case of $t_{2g}$ model 
[\fref{fig:chiEz2}(b)] one finds the AF character of $\chi_s$ in the 
interesting range of $J_H$. Altogether, as for $E_z=0$, the 
spin susceptibility $\chi_s$ exhibits again the opposite behavior for 
both types of orbitals --- the high-spin state is suppressed in the 
$t_{2g}$ model, while it can be selected by the crystal field in the 
$e_g$ model. The maximum of $\chi_s$ occurs at higher temperature in 
$t_{2g}$ than in $e_g$ model at $J_H=U/8$. 

\subsection{Specific heat and entropy at finite crystal field}

\begin{figure}[t!]
\begin{center}
\includegraphics*[width=0.75\textwidth ]{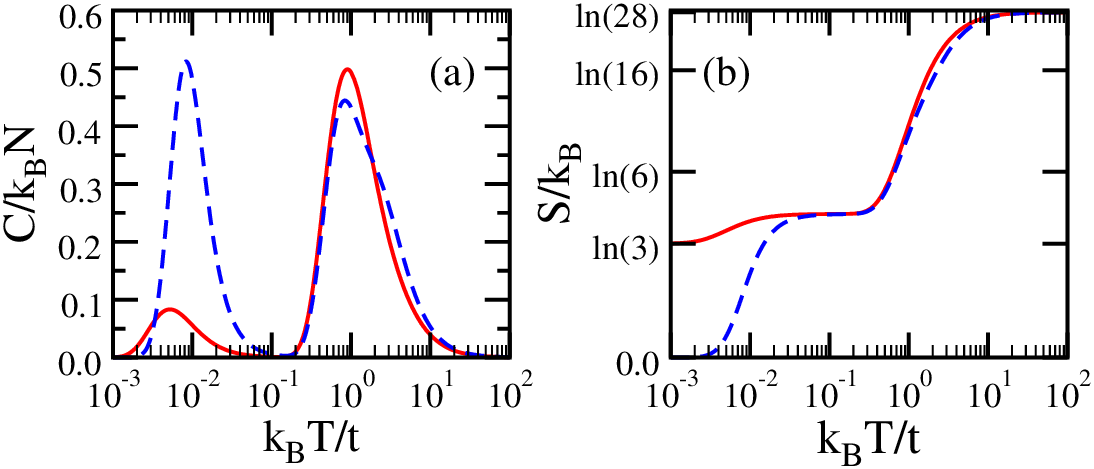}
\end{center}
\caption
{ Temperature dependence of:
(a) the specific heat $C$, and (b) the entropy $S$ for $e_g$ orbitals. 
Dashed (solid) lines corresponds to $J_H=U/8$ ($J_H=U/4$), respectively.
Parameters: $U=8t$ and $E_0=2t$. 
}
\label{fig:SEz2eg}
\end{figure}

As expected from the above results, a finite crystal field also modifies 
the temperature evolution of the specific heat $C$ and the entropy $S$
for $e_g$ orbitals, see figure \fref{fig:SEz2eg}. The position of the 
high temperature peak of the specific heat at $k_BT\simeq t$ is almost 
the same as the position of the strong anomalies of both on-site 
correlation functions (\textit{cf}. figures \ref{fig:SEz2eg}(a) and 
\ref{fig:Ez2eg}). Depending on $J_H$, the low temperature entropy $S$ of 
the $e_g$ system either vanishes (for the singlet ground state as 
$J_H\le U/8$) or approaches the value $k_B\ln 3$ (for the triplet ground 
state at $J_H=U/4$). Nevertheless, owing to the vanishing specific heat, 
all the curves in \fref{fig:SEz2eg}(b) have a point of inflection 
$S=k_B\ln 4$ at $k_BT\simeq 0.1t$. Note that in contrast to the case 
without a crystal field, there is no such a point when $S=k_B\ln16$. 
Namely, by promoting one pseudospin over the other one, a finite crystal 
field markedly lowers the states with double occupancies, hence the 
usual gap between the singly and doubly occupied states vanishes.

\begin{figure}[t!]
\begin{center}
\includegraphics*[width=0.75\textwidth ]{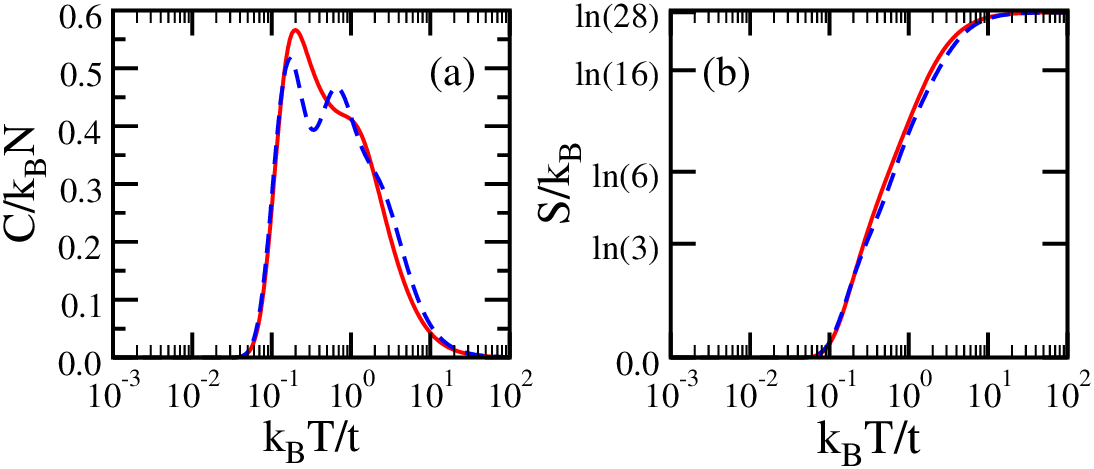}
\end{center}
\caption
{ 
The same as in \fref{fig:SEz2eg} but for $t_{2g}$ orbitals. 
Dashed (solid) line corresponds to $J_H=U/8$ ($J_H=U/4$), respectively. 
}
\label{fig:SEz2t2g}
\end{figure}

Consider now the temperature behaviour of the specific heat for $t_{2g}$ 
orbitals shown in \fref{fig:SEz2t2g}(a). Instead of the high and low 
temperature peaks of the specific heat one observes two slightly 
separated peaks for $J_H=U/8$ which merge into a wide peak at $J_H=U/4$.
This confirms that spin-orbital degrees of freedom are strongly coupled
as spin and orbital intersite correlations change at the same 
temperature (see \fref{fig:Ez2t2g}). 
Finally, as shown in \fref{fig:SEz2t2g}(b), the entropy of the $t_{2g}$
system is almost independent of $J_H$. We note that $S$ is entirely
suppressed at low $T$ due to a spin singlet ground state; it rises 
rapidly at $k_BT\simeq 0.1t$, and approaches eventually the limiting 
value $S=k_B\ln 28$, not having any point of inflection. Such a 
behaviour is a direct consequence of a single broad peak in the specific 
heat and no separation of the energy scales for spin and orbital 
excitations.

\begin{figure}[t!]
\begin{center}
\includegraphics*[width=0.75\textwidth ]{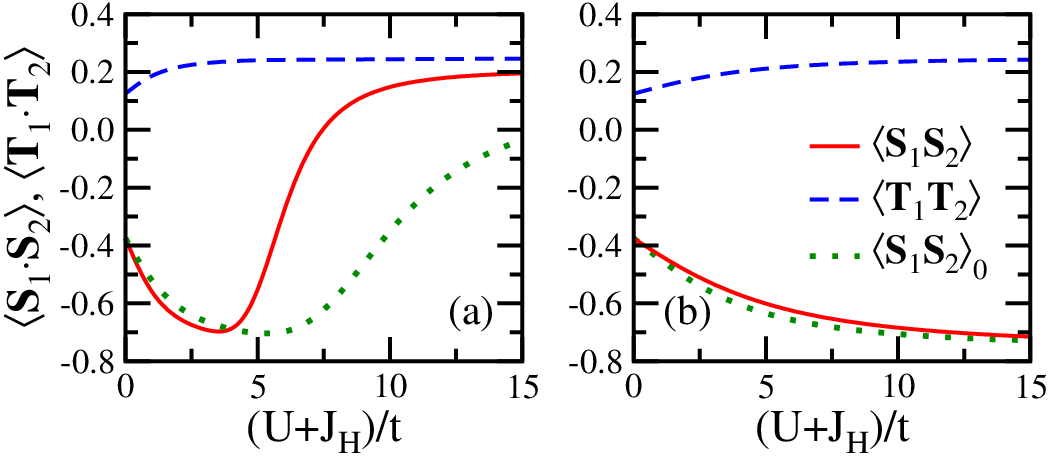}
\end{center}
\caption
{Zero temperature intersite correlation functions: 
spin $\langle{\textbf S}_1\cdot{\textbf S}_{2}\rangle$ (solid lines) 
and pseudospin $\langle{\textbf T}_1\cdot{\textbf T}_{2}\rangle$ 
(dashed lines) as functions of the Stoner parameter $U+J_H$ for: 
(a) $e_g$ orbitals, and 
(b) $t_{2g}$ orbitals. 
Parameters: $E_0=2t$, $J_H=U/4$.  
Dotted line shows the spin-spin correlation function 
$\langle{\textbf S}_1\cdot{\textbf S}_{2}\rangle_0$ 
obtained in the Ising limit (see text).
}
\label{fig:fUJHEz2}
\end{figure}

We close this section with a short discussion of the intersite spin
$\langle{\textbf S}_1\cdot{\textbf S}_2\rangle$ and pseudospin 
$\langle{\textbf T}_1\cdot{\textbf T}_2\rangle$ correlations for the 
$e_g$ system depicted in \fref{fig:fUJHEz2}(a) as functions of the 
Stoner parameter, $I=U+J_H$, in the presence of a finite crystal field 
$E_0=2t$. As expected, the results illustrate the AF correlation between 
spins on two sites in the weak coupling regime $U+J_H\lesssim 7t$.
However, further increase of the interaction strength changes gradually 
the AF coupling into a FM one, with the immobile $|\xi\rangle$ 
pseudospin component preferred by the crystal field. Remarkably, such a 
transition at $U+J_H\simeq 16t$ in the presence of the same crystal 
field $E_0=2t$ from the AF phase into the FM one has been obtained on
the infinite lattice in the Hartree approximation recently \cite{Rac05}. 
(The MF phase diagram appears more involved \cite{Fre05}, since FM 
order is stabilized for $5 \le I/t \le 6.5$, in contrast to what is
obtained in the present study. Nevertheless this mostly reflects the
strong competition between the FM and AF phases, since their energy do
not differ by more than $1$\% in this regime \cite{Rac06b}.) 
Consequently, owing to 
a strong competition between the singlet and triplet states with the 
lowest energies (\textit{cf}. \tref{tab:EDEz2eg}), respectively, one 
can conjecture that fluctuations clearly affect the intersite spin 
correlations. Indeed, comparison of  
$\langle{\textbf S}_1\cdot{\textbf S}_2\rangle$ and 
$\langle{\textbf S}_1\cdot{\textbf S}_2\rangle_0$ determined in the 
Ising limit (i.e., taking only $S_i^z$ operators and neglecting the
last term in the Hamiltonian (\ref{eq:H_int})) shows that 
dynamics acts to reduce the AF coupling between spins. In contrast, 
finite crystal field affects 
$\langle{\textbf T}_1\cdot{\textbf T}_{2}\rangle$ only slightly, and 
the pseudospin correlations are always positive for increasing $I$ 
regardless of $E_0$, which implies the FO coupling between two 
pseudospins [\textit{cf}. figures \ref{fig:fUJHEz0}(a) and 
\ref{fig:fUJHEz2}(a)]. Thus, in the regime of large Stoner parameter $I$
one finds that the classical Goodenough-Kanamori rules \cite{Goode} are 
violated as the crystal field stabilizes a particular orbital 
configuration. In contrast to quantum spin-orbital entanglement 
\cite{Ole06}, this effect here is static and may play an important role 
for the observed magnetic and orbital correlations in transition metal 
oxides, as discussed recently on the example of LiNiO$_2$ \cite{Hay03}.

Turning now to the $t_{2g}$ model with positive 
$\langle{\textbf S}_1\cdot{\textbf S}_{2}\rangle$ and negative 
$\langle{\textbf T}_1\cdot{\textbf T}_{2}\rangle$ in the degenerate case 
shown in \fref{fig:fUJHEz0}(b), the situation is also changed 
drastically by a finite crystal field as depicted in 
\fref{fig:fUJHEz2}(b). Indeed, the resulting 
$\langle{\textbf S}_1\cdot{\textbf S}_{2}\rangle$ is then negative 
revealing the AF nature of the ground state. However, due to the energy 
gap between a singlet ground state and the first excited triplet state 
(\textit{cf}. \tref{tab:EDEz2t2g}), fluctuations modify the value of the 
intersite spin correlations only slightly, so that 
$\langle{\textbf S}_1\cdot{\textbf S}_{2}\rangle$ and 
$\langle{\textbf S}_1\cdot{\textbf S}_{2}\rangle_0$ almost overlap.
Finally, positive $\langle{\textbf T}_1\cdot{\textbf T}_{2}\rangle$ 
shows that the AO pseudospin correlations found before at $E_0=0$ are 
suppressed. 

\section{Summary}
\label{sec:sum}

We have obtained the exact numerical results for a two-site Hubbard 
model with two $e_g$ orbitals at quarter-filling. By an appropriate 
transformation of the original orbital basis $\{|x\rangle,|z\rangle\}$,
into a basis consisting of a directional orbital $|\zeta\rangle$ along 
the molecular bond and a planar and orthogonal to it orbital 
$|\xi\rangle$, we have simplified the hopping matrix making use of the 
properties of $e_g$ orbitals --- the electrons in $|\zeta\rangle$ 
directional orbitals are mobile, while those in orthogonal $|\xi\rangle$ 
orbitals are fully localized \cite{Fei05}. The results were 
compared with the doubly degenerate Hubbard model with two equivalent 
$t_{2g}$ orbitals active along the molecular bond direction.  

As the first important result, a striking difference between $t_{2g}$ 
and $e_{g}$ orbitals with respect to the ground state has been 
established. Indeed, in contrast to the $t_{2g}$ model, where even 
infinitesimally small Hund's exchange $J_H>0$ lifts the degeneracy of
the lowest energy singlet and triplet states and stabilizes the FM spin 
correlation, we have found that a spin singlet ground state survives up 
to $J_H\simeq 0.27U$ for $e_g$ electrons. In this regime of parameters, 
the intersite spin correlations 
$\langle{\textbf S}_1\cdot{\textbf S}_2\rangle$ 
indicate the AF nature of the ground state, whereas the pseudospin 
function $\langle{\textbf T}_1\cdot{\textbf T}_2\rangle$ illustrates 
the FO pseudospin correlation. On the contrary, in the 
$t_{2g}$ model, except for $J_H=0$ when the intersite spin and 
pseudospin correlation functions overlap and are negative, positive 
$\langle{\textbf S}_1\cdot{\textbf S}_2\rangle$ demonstrates the FM 
nature of the ground state supported by the pseudospin singlet with 
negative $\langle{\textbf T}_1\cdot{\textbf T}_2\rangle$, i.e. AO 
correlations. Therefore, a complementary behaviour of the spin and 
orbital flavours is observed as a generic feature of both models, 
in agreement with the Goodenough-Kanamori rules \cite{Goode}.

We have further demonstrated the influence of a finite crystal field 
$E_0$ on the ground state. On the one hand, it suppresses the FM spin 
correlation in the $t_{2g}$ model and stabilizes the spin singlet 
ground state with negative  
$\langle{\textbf S}_1\cdot{\textbf S}_{2}\rangle$, accompanied by 
positive $\langle{\textbf T}_1\cdot{\textbf T}_{2}\rangle$. On the other 
hand, the effect of the crystal field on the $e_g$ ground state is  
the opposite one. Namely, by lifting the degeneracy of the pseudospin 
flavours it promotes the immobile $|\xi\rangle$ one. Consequently, not 
much kinetic energy can be gained and the Coulomb interactions start to 
dominate. However, they are noticeably better optimized by the FM spin 
correlation. Therefore, it becomes energetically advantageous to have 
the spin triplet and the ground state for large $J_H=U/4$ yields 
positive $\langle{\textbf S}_1\cdot{\textbf S}_2\rangle$, while a 
smaller Hund's exchange coupling $J_H=U/8$ drives the system towards 
the singlet in the ground state which results in negative 
$\langle{\textbf S}_1\cdot{\textbf S}_2\rangle$. In contrast,
$\langle{\textbf T}_1\cdot{\textbf T}_2\rangle$ is positive and almost 
insensitive to the actual value of $J_H$ in the range $0<J_H<U/4$.  

Next, we have investigated how the $e_g$ ground state properties evolve 
as a function of the Stoner parameter $I=U+J_H$ in the presence of a 
finite crystal field $E_0=2t$. As expected, we have found the AF 
correlation between spins on two sites in the weak coupling regime  
$U+J_H\lesssim 7t$. However, further increase of the interaction 
strength changes gradually the AF coupling into a FM one, with an
immobile $|\xi\rangle$ pseudospin, preferred by the crystal field. 
Finally, by comparing the correlation function 
$\langle{\textbf S}_1\cdot{\textbf S}_{2}\rangle$ with the one 
determined in the Ising limit 
$\langle{\textbf S}_1\cdot{\textbf S}_{2}\rangle_0$, we have elucidated 
the role of fluctuations in the intersite spin correlation function and 
have observed that the dynamics helps to reduce strongly the AF 
coupling between spins in this case. In contrast, when the ground state 
is FM, fluctuations only slightly modify the value of the intersite spin 
correlations, so that 
$\langle{\textbf S}_1\cdot{\textbf S}_{2}\rangle$ and 
$\langle{\textbf S}_1\cdot{\textbf S}_{2}\rangle_0$ almost overlap.

Although our results presented here are only a starting point 
and a systematic analysis of the properties of the ground states and 
spin-orbital correlations in realistic two-band models with $e_g$ and 
$t_{2g}$ orbitals of higher dimension is required, our study shows that 
the description of transition metal oxides with partly filled (almost)
degenerate orbitals has to involve correct symmetry of the orbital
degrees of freedom, and thus has to go beyond the simplified Hubbard
model with degenerate equivalent orbitals. This observation is also 
supported by our recent study of stripe phases in the nickelates 
\cite{Rac06}. Indeed, the diagonal stripe structures with filling of 
nearly one hole per atom, as observed experimentally, are the ground 
state of the model with the physically relevant hopping elements 
between $e_g$ orbitals, while instead the most stable stripes are 
half-filled within the doubly degenerate Hubbard model. We therefore 
conclude that the ground state properties strongly depend on the 
orbital degrees of freedom active in a given compound.

\ack 
We thank C. Lacroix and V. Caignaert for valuable and encouraging 
discussions. M. Raczkowski acknowledges support from the European 
Community under Marie Curie Program number HPMT2000-141. This work was 
supported by the the Polish Ministry of Science and Education, Project 
No. 1~P03B~068~26, and by the Minist\`ere Fran\c{c}ais des Affaires 
Etrang\`eres under POLONIUM 09294VH.

\appendix

\section{Construction of the basis for the two-site model}
\label{ap:states}

We construct a basis of the Hilbert space, starting with the $S^z=1$ 
subspace. There are two states with $T^{\zeta}=\pm 1$,
\begin{eqnarray}
|\Psi_{\zeta\uparrow}^{}\rangle =
c_{1\zeta\uparrow}^{\dag}c_{2\zeta\uparrow}^{\dag}|0\rangle,  \qquad
|\Psi_{\xi\uparrow}^{}\rangle   =
c_{1\xi\uparrow}^{\dag}c_{2\xi\uparrow}^{\dag}|0\rangle,
\end{eqnarray}
and four with $T^{\zeta}=0$,
\begin{eqnarray}
|\Psi_{1\uparrow}^{\pm}\rangle &=\case{1}{\sqrt{2}}\bigl(
c_{1\xi\uparrow}^{\dag}c_{2\zeta\uparrow}^{\dag} \pm
c_{1\zeta\uparrow}^{\dag}c_{2\xi\uparrow}^{\dag}\bigr)|0\rangle, \\
|\Psi_{2\uparrow}^{\pm}\rangle &=\case{1}{\sqrt{2}}\bigl(
c_{1\xi\uparrow}^{\dag}c_{1\zeta\uparrow}^{\dag} \pm
c_{2\xi\uparrow}^{\dag}c_{2\zeta\uparrow}^{\dag}\bigr)|0\rangle.
\end{eqnarray}
In the $S^z=0$ subspace there are eight states with $T^{\zeta}=\pm 1$,
\begin{eqnarray}
|\Phi_1^{\pm}\rangle &=\case{1}{\sqrt{2}}\bigl(
c_{1\zeta\uparrow}^{\dag}c_{2\zeta\downarrow}^{\dag}   \pm
c_{1\zeta\downarrow}^{\dag}c_{2\zeta\uparrow}^{\dag}\bigr)|0\rangle, \\
|\Phi_2^{\pm}\rangle &=\case{1}{\sqrt{2}}\bigl(
c_{1\zeta\uparrow}^{\dag}c_{1\zeta\downarrow}^{\dag}   \pm
c_{2\zeta\uparrow}^{\dag}c_{2\zeta\downarrow}^{\dag}\bigr)|0\rangle, \\
|\Phi_3^{\pm}\rangle &=\case{1}{\sqrt{2}}\bigl(
c_{1\xi\uparrow}^{\dag}c_{1\xi\downarrow}^{\dag}  \pm
c_{2\xi\uparrow}^{\dag}c_{2\xi\downarrow}^{\dag}\bigr)|0\rangle,    \\
|\Phi_4^{\pm}\rangle &=\case{1}{\sqrt{2}}\bigl(
c_{1\xi\uparrow}^{\dag}c_{2\xi\downarrow}^{\dag}   \pm
c_{1\xi\downarrow}^{\dag}c_{2\xi\uparrow}^{\dag}\bigr)|0\rangle,
\end{eqnarray}
and eight with $T^{\zeta}=0$,
\begin{eqnarray}
|\Phi_5^{\pm}\rangle &=\case{1}{2}\Bigl(
c_{1\xi\uparrow}^{\dag}c_{1\zeta\downarrow}^{\dag}   +
c_{1\xi\downarrow}^{\dag}c_{1\zeta\uparrow}^{\dag}   \pm \bigl(
c_{2\xi\uparrow}^{\dag}c_{2\zeta\downarrow}^{\dag}+
c_{2\xi\downarrow}^{\dag}c_{2\zeta\uparrow}^{\dag}\bigr)\Bigr)|0\rangle, \\
|\Phi_6^{\pm}\rangle &=\case{1}{2}\Bigl(
c_{1\xi\uparrow}^{\dag}c_{2\zeta\downarrow}^{\dag}  +
c_{1\xi\downarrow}^{\dag}c_{2\zeta\uparrow}^{\dag}  \pm \bigl(
c_{1\zeta\downarrow}^{\dag}c_{2\xi\uparrow}^{\dag}  +
c_{1\zeta\uparrow}^{\dag}c_{2\xi\downarrow}^{\dag}\bigr)\Bigr)|0\rangle, \\
|\Phi_7^{\pm}\rangle &=\case{1}{2}\Bigl(
c_{1\xi\uparrow}^{\dag}c_{1\zeta\downarrow}^{\dag}   -
c_{1\xi\downarrow}^{\dag}c_{1\zeta\uparrow}^{\dag}   \pm \bigl(
c_{2\xi\uparrow}^{\dag}c_{2\zeta\downarrow}^{\dag}   -
c_{2\xi\downarrow}^{\dag}c_{2\zeta\uparrow}^{\dag}\bigr)\Bigr)|0\rangle, \\
|\Phi_8^\pm\rangle &=\case{1}{2}\Bigl(
c_{1\xi\uparrow}^{\dag}c_{2\zeta\downarrow}^{\dag}  -
c_{1\xi\downarrow}^{\dag}c_{2\zeta\uparrow}^{\dag}  \pm \bigl(
c_{1\zeta\downarrow}^{\dag}c_{2\xi\uparrow}^{\dag}  -
c_{1\zeta\uparrow}^{\dag}c_{2\xi\downarrow}^{\dag}\bigr)\Bigr)|0
\rangle.
\end{eqnarray}
The states $|\Phi_1^{+}\rangle $ and $|\Phi_4^{+}\rangle $, together
with $|\Phi_5^{\pm}\rangle$ and $|\Phi_6^{\pm}\rangle$ belong
to the triplet subspace.

\section{Eigenvalues of the Hamiltonian (\ref{eq:H_deg}) with  
$t_{2g}$ orbitals}
\label{ap:t2g}

Below we give a complete list of the eigenvalues of Hamiltonian 
(\ref{eq:H_deg}) for $t_{2g}$ orbitals in the absence of crystal 
field splitting (at $E_0=0$) with the degeneracy of the states 
given in parenthesis:

--- $S=1$ subspace

\begin{equation}
\case{1}{2}\Big\{U-3J_H\pm\sqrt{(U-3J_H)^2 + 16t^2}\Big\}\;(3), 
\hskip .3cm  U-3J_H\;(3), \hskip .3cm 0\;(9)\;;
\label{eq:T_t2g}
\end{equation}

--- $S=0$ subspace
\begin{equation}
\case{1}{2} \Big\{U^{} - J_H\pm\sqrt{(U -J_H)^2 + 16t^2}\Big\}\;(2), 
\qquad U-J_H\;(2)\;,
\label{eq:S_t2ga}
\end{equation}
and
\begin{equation}
\case{1}{2} \Big\{U^{} + J_H\pm\sqrt{(U +J_H)^2 + 16t^2}\Big\},
\qquad  U+J_H,    \qquad 0\;.
\label{eq:S_t2gb}
\end{equation}

\section{Eigenvalues of the Hamiltonian (\ref{eq:H_deg}) with  
$e_{g}$ orbitals}
\label{ap:eg}

Finally we present the eigenvalues of the Hamiltonian given by equation
\eref{eq:H_deg} for $e_{g}$ orbitals in the absence of crystal field 
splitting (at $E_0=0$). 
The degeneracy of the states is given in parenthesis:

--- $S=1$ subspace

\begin{equation}
\case{1}{2}\Big\{U-3J_H\pm\sqrt{(U-3J_H)^2 + 4t^2}\Big\} \;(6), 
\qquad 0\;(6)\;;
\label{eq:T_eg}
\end{equation}

--- $S=0$ subspace

\begin{equation}
\case{1}{2} \Big\{U^{} - J_H\pm\sqrt{(U -J_H)^2 + 4t^2}\Big\}\;(2), 
\qquad  U\pm J_H,    \qquad  0\;,
\label{eq:S_ega}
\end{equation}
and,
\begin{eqnarray}
\label{eq:Seg1}
\lambda_{-1} &=& \frac{2U}{3}\Big\{1-\sqrt{1+3\Bigl(\frac{J_H}{U}\Bigr)^2 
 + 12\Bigl(\frac{t}{U}\Bigr)^2}\cos{\Bigl(\frac{\alpha}{3}\Bigr)}\Big\},\\
\label{eq:Seg2}
\lambda_{0} &=& \frac{2U}{3} \Big\{1+\sqrt{1+3\Bigl(\frac{J_H}{U}\Bigr)^2 
+12\Bigl(\frac{t}{U}\Bigr)^2}\cos{\Bigl(\frac{\pi+\alpha}{3}\Bigr)}\Big\},\\
\label{eq:Seg3}
\lambda_{1} &=& \frac{2U}{3} \Big\{1+\sqrt{1+3\Bigl(\frac{J_H}{U}\Bigr)^2 
+12\Bigl(\frac{t}{U}\Bigr)^2}\cos{\Bigl(\frac{\pi-\alpha}{3}\Bigr)}\Big\}, 
\end{eqnarray}
with
\begin{equation}
\cos{(\alpha)}= 
\frac{1-\big(\frac{3J_H}{U}\big)^2  + 2\big(\frac{3t}{U}\big)^2}
{\Big\{1+3\big(\frac{J_H}{U}\big)^2 +12\big(\frac{t}{U}\big)^2\Big\}
^{\frac{3}{2}}}.
\label{eq:angle}
\end{equation}
The latter three eigenvalues follow from the submatrix in equation 
\eref{eq:S3}.


\end{document}